\documentclass[apj,twocolappendix]{emulateapj}
\usepackage{amssymb}
\usepackage{amsmath}
\usepackage{cancel}
\usepackage{bbold}
\usepackage{graphicx}
\usepackage{xspace}
\usepackage{units}
\usepackage[breaklinks,colorlinks,urlcolor=blue,citecolor=blue,linkcolor=blue]{hyperref}

\let\oldsqrt\sqrt
\def\sqrt{\mathpalette\DHLhksqrt}
\def\DHLhksqrt#1#2{
	\setbox0=\hbox{$#1\oldsqrt{#2\,}$}
	\dimen0=\ht0
	\advance\dimen0-0.2\ht0
	\setbox2=\hbox{\vrule height\ht0 depth -\dimen0}
	{\box0\lower0.4pt\box2}
}

\newcommand{\bo}{\raise-0.4pt\hbox{\large$\Box$}}			
\renewcommand{\d}{\ensuremath{\mathrm{d}}}					
\renewcommand{\arg}[1]{\ensuremath{\! \left( #1 \right)}}

\newcommand{\pc}{\ensuremath{\, \mathrm{pc}}\xspace}
\newcommand{\kpc}{\ensuremath{\, \mathrm{kpc}}\xspace}

\newcommand{\FeH}{\ensuremath{\left[ \nicefrac{\mathrm{Fe}}{\mathrm{H}} \right]}}
\newcommand{\EBV}{\ensuremath{\mathrm{E} \! \left( B \! - \! V \right)}}

\begin{document}

\title{A Three-Dimensional Map of Milky Way Dust}

\author{Gregory~M.~Green\altaffilmark{1},
Edward~F.~Schlafly\altaffilmark{2},
Douglas~P.~Finkbeiner\altaffilmark{1},
Hans-Walter~Rix\altaffilmark{2},
Nicolas~Martin\altaffilmark{3,2},
William~Burgett\altaffilmark{4},
Peter~W.~Draper\altaffilmark{5},
Heather~Flewelling\altaffilmark{6},
Klaus~Hodapp\altaffilmark{6},
Nicholas~Kaiser\altaffilmark{6},
Rolf~Peter~Kudritzki\altaffilmark{6},
Eugene~Magnier\altaffilmark{6},
Nigel~Metcalfe\altaffilmark{5},
Paul~Price\altaffilmark{7},
John~Tonry\altaffilmark{6},
Richard~Wainscoat\altaffilmark{6}
}

\email{ggreen@cfa.harvard.edu}

\altaffiltext{1}{Harvard-Smithsonian Center for Astrophysics, 60 Garden St., Cambridge, MA 02138}
\altaffiltext{2}{Max-Planck-Institut f\"{u}r Astronomie, K\"{o}nigstuhl 17, D-69117 Heidelberg}
\altaffiltext{3}{Observatoire astronomique de Strasbourg, Universit\'e de Strasbourg, CNRS, UMR 7550, 11 rue de l'Universit\'e, F-67000 Strasbourg, France}
\altaffiltext{4}{GMTO Corporation, 251 S. Lake Ave., Suite 300, Pasadena, CA 91101, USA}
\altaffiltext{5}{Department of Physics, University of Durham, South Road, Durham DH1 3LE, UK}
\altaffiltext{6}{Institute for Astronomy, University of Hawaii at Manoa, 2680 Woodlawn Dr., Honolulu, HI 96822, USA}
\altaffiltext{7}{Princeton University Observatory, 4 Ivy Lane, Peyton Hall, Princeton University, Princeton, NJ 08544, USA}

\begin{abstract}
	We present a three-dimensional map of interstellar dust reddening, covering three-quarters of the sky out to a distance of several kiloparsecs, based on Pan-STARRS 1 and 2MASS photometry. The map reveals a wealth of detailed structure, from filaments to large cloud complexes. The map has a hybrid angular resolution, with most of the map at an angular resolution of $3.4^{\prime}$ to $13.7^{\prime}$, and a maximum distance resolution of $\sim \! 25\%$. The three-dimensional distribution of dust is determined in a fully probabilistic framework, yielding the uncertainty in the reddening distribution along each line of sight, as well as stellar distances, reddenings and classifications for 800 million stars detected by Pan-STARRS 1. We demonstrate the consistency of our reddening estimates with those of two-dimensional emission-based maps of dust reddening. In particular, we find agreement with the Planck $\tau_{353 \, \mathrm{GHz}}$-based reddening map to within $0.05 \, \mathrm{mag}$ in $\EBV$ to a depth of $0.5 \, \mathrm{mag}$, and explore systematics at reddenings less than $\EBV \approx 0.08 \, \mathrm{mag}$. We validate our per-star reddening estimates by comparison with reddening estimates for stars with both SDSS photometry and SEGUE spectral classifications, finding per-star agreement to within $0.1 \, \mathrm{mag}$ out to a stellar $\EBV$ of $1 \, \mathrm{mag}$. We compare our map to two existing three-dimensional dust maps, by \citet{Marshall2006a} and \citet{Lallement2013}, demonstrating our finer angular resolution, and better distance resolution compared to the former within $\sim \! 3 \kpc$. The map can be queried or downloaded at \url{http://argonaut.skymaps.info}. We expect the three-dimensional reddening map presented here to find a wide range of uses, among them correcting for reddening and extinction for objects embedded in the plane of the Galaxy, studies of Galactic structure, calibration of future emission-based dust maps and determining distances to objects of known reddening.
\end{abstract}

\section{Introduction}
\label{sec:introduction}

The Milky Way is the only Galaxy that can be observed in such close detail, yet most of the plane of the Galaxy is veiled by dust. While dust makes up just 1\% of the mass of the interstellar medium, it absorbs on the order of 30\% of all starlight \citep{Draine2003}. Interstellar dust absorbs and scatters light in the ultraviolet (UV), optical and near-infrared (NIR), re-radiating it in the mid- to far-infrared (MIR and FIR, respectively). Understanding the spatial distribution of dust is therefore critical for UV, optical and NIR astronomy, where dust is an extinguishing foreground, to extragalactic astronomy and cosmology, where it is a radiating foreground, as well as to star formation, where the dust itself is a primary object of study. Detailed studies of stellar populations and substructures in the Galaxy require accurate corrections for extinction and reddening due to dust. The plane of the Milky Way, containing the majority of the Galaxy's stellar content, is also the most heavily extinguished region. But dust is not only a nuisance to astronomers. Dust traces the interstellar medium, and a clearer picture of the spatial distribution of dust would aid in understanding the processes that shape the Galaxy, from star formation to how feedback from supernovae and stellar winds shape our Galaxy's ISM.

Dust can be mapped either through its extinction or through its emission. The spectral energy distribution and amplitude of FIR dust emission is most sensitive to dust column density, temperature, and grain-size distribution. By modeling these properties across the sky, a map of dust column density may be obtained, which can then be converted to extinction or reddening by calibration against astrophysical extinction or reddening standards. Such emission-based methods recover the angular distribution of dust, but not its distribution in distance. \citet{Burstein1982} produced an all-sky map of dust reddening based on H\textsc{i} emission and galaxy counts, assuming the gas and dust to be uniformly mixed. \citet{Schlegel1998} (hereafter ``SFD'') produced a widely-used all-sky reddening map, using DIRBE and IRAS maps of FIR emission to model dust column density and temperature. More recently, the Planck Collaboration has released all-sky reddening maps based on a similar emission-modeling technique \citep{PlanckCollaboration2013}.

A second class of dust maps is based on extinction or reddening estimates of sources distributed across the sky. \citet[][NICE]{Lada1994} compared the average $H - K$ colors and number counts of stars in target and control fields to map dust reddening in two dimensions. \citet[][NICER]{Lombardi2001} and \citet[][NICEST]{Lombardi2009} extended on this algorithm, which has now been applied to 2MASS photometry to obtain reddening maps of a host of cloud complexes \citep{Lombardi2006,Lombardi2008,Lombardi2010,Lombardi2011,Alves2014}. Froebrich applied a similar \textit{color excess} method to 2MASS photometry to produce a two-dimensional, all-sky reddening map \citep{Rowles2009}. \citet{Schlafly2010} used the location of main-sequence turn-off stars in color-color space to simultaneously map dust reddening and test different reddening laws. \citet{Peek2010} used passively evolving galaxies as a standard color source in order to map reddening and correct the SFD map.

Because stars are distributed throughout the Galaxy, they can be used to trace the dust distribution in three dimensions. Most methods relying on this principle group stars into separate sightlines, and then determine stellar reddening as a function of distance along each sightline. The challenge with such methods is to simultaneously determine stellar type (and thus the intrisic stellar colors and luminosity), distance and reddening on the basis of photometry alone. \citet{Marshall2006a} developed a method that iteratively improves distance and reddening estimates to post-main sequence stars, updating the dust column in each distance bin with each iteration so that the intrinsic stellar colors match those predicted by the Besan\c{c}on model of the Galactic stellar population \citep{Robin2003}. \citet{Marshall2006a} applied this method to 2MASS photometry of the Galactic plane, producing a three-dimensional reddening map out to a distance of several kiloparsecs in the region $\left| b \right| < 10^{\circ}$, $-100^{\circ} \leq \ell \leq 100^{\circ}$. A number of groups have pursued methods that determine maximum-likelihood parameters for the stars along each sightline, combining the estimates in each sightline into a distance-reddening relation. \citet{Berry2011} applied such a method to SDSS photometry, producing a 3D reddening map and measuring variation in the dust extinction spectrum across the SDSS footprint. \citet{Chen2014} used optical XSTPS-GAC, NIR 2MASS and MIR WISE photometry to produce a 3D reddening map of the Galactic anticenter. \citet{Sale2012} developed a probabilistic framework to simultaneously infer stellar parameters and the dust extinction distribution along each line of sight, and \citet{Sale2014b} applied this method to IPHAS photometry of the northern Galactic plane. \citet{Hanson2014} developed a similar probabilistic framework for inferring stellar types, distances and dust foreground properties, and applied this method to SDSS and UKIRT Infrared Deep Sky Survey photometry in regions of the high-Galactic-latitude sky. \citet{Lallement2013} took a somewhat different approach, using $\sim 23,000$ stellar parallaxes and reddening estimates from a number of sources to infer the 3D distribution of dust opacity out to a distance of $800$ - $1000 \, \mathrm{pc}$ in the plane of the Galaxy, and $\sim 300 \, \mathrm{pc}$ out of the plane. \citet{Green2014} developed a probabilistic method for determining dust reddening in 3D from stellar photometry, and \citet{Schlafly20142d} presented a map of dust reddening integrated to 4.5\kpc, covering three quarters of the sky, based on applying this method to optical Pan-STARRS 1 stellar photometry. Using this same 3D mapping technique, \citet{Schlafly2014Orion} found that the Orion molecular complex appears to form part of a larger bubble structure. \citet{Schlafly2014clouds} determined the distance to a large number of molecular clouds by applying a related method to Pan-STARRS 1 stellar photometry.

In this paper, we present a three-dimensional map of dust reddening over three quarters of the sky. Our map covers the Northern sky down to $\delta \approx -30^{\circ}$, has an adaptive resolution of typical angular scale $3.4^{\prime}$ to $13.7^{\prime}$, logarithmically-spaced distance bins with a width of $\sim \! 25\%$, and extends to $\EBV \sim 1.5 - 2 \, \mathrm{mag}$. Our map is based on high-quality Pan-STARRS 1 (PS1) photometry for 800 million stars, $\sim \! 200$ million of which have matched 2MASS photometry. The reddening along each sightline is determined in a fully probabilistic manner, according to the method developed in \citet{Green2014}. We strongly encourage readers who wish to understand our method for inferring the three-dimensional distribution of dust reddening to read that paper, as the present paper focuses primarily on results obtained using that method.

This paper is organized as follows. In \S\ref{sec:method}, we briefly describe the method developed in \citet{Green2014}, and describe the priors we place on the three-dimensional distribution of dust reddening. In \S\ref{sec:data}, we describe the data sources which go into our map. In \S\ref{sec:results}, we discuss results from the map, and in \S\ref{sec:comparison}, we compare our map to existing 3D and 2D dust maps. We describe how the map can be accessed in \S\ref{sec:accessing}.

\section{Method}
\label{sec:method}

Our three-dimensional dust map is constructed pixel-by-pixel from independently determined line-of-sight reddening profiles. We group stars into sightlines with typical angular scales of $\sim 6.8^{\prime}$, probabilistically inferring the reddening, distance and stellar type of each star independently. We then assume that the distances and reddenings of the stars in a given line of sight lie along a single profile, with reddening increasing monotonically with distance. We sample from the posterior density of distance-reddening profiles, returning the uncertainty in the line-of-sight reddening in the pixel. By repeating this process in each angular pixel, we obtain the 3D distribution of dust throughout the Pan-STARRS 1 survey volume \citep[PS1;][]{PS1_system}. The details of our method are described in \citet{Green2014}.

We have made three notable modifications to our method since publication of \citet{Green2014}. The first modification is the inclusion of stellar photometry from the Two Micron All Sky Survey (2MASS). We compile joint PS1+2MASS stellar templates, in a manner very similar to the process laid out for the PS1 templates used in \citet{Green2014}. Appendix \ref{app:PS1-2MASS-models} contains details about the construction of our PS1+2MASS templates, while Appendix \ref{app:2MASS-selection} details our estimate of the 2MASS survey depth.

The second modification to our method is to allow the reddening of individual stars to deviate slightly from the overall line-of-sight reddening profile in a pixel. Our basic model assumes that within an angular pixel, the dust density does not depend on angle, but is solely a function of distance. This is a good assumption at very fine angular scales, but it must obviously begin to break down at some angular scale, depending on the power spectrum of the dust density field. Because each star in an angular pixel is behind a slightly different column of dust, we allow the total column of dust in front of each star to vary slightly from the modeled column density. For a pixel with a small angular scale, in which the dust column should be more uniform across the pixel, the scatter of the stars around the modeled distance-reddening profile should be small. For pixels with larger angular scale, where angular variations in the dust column are correpsondingly greater, one should allow the stars to deviate from the modeled distance-reddening profile to a greater extent. We therefore include an additional parameter for each star, describing the fractional variation of the star from the modeled distance-reddening profile in its pixel. This modification to the model amounts to introducing a smoothing factor in the probability density of each star in distance-reddening space, and therefore contributes only negligibly to the overall computation time. We discuss this modification in more detail in\S\ref{sec:los-scatter}.

A third modification we have made is to adopt the increased, bias-corrected stellar disk scale lengths and heights presented in Table 10 of \citet{Juric2008} for our disk priors. We also adopt a less dense halo prior than \citet{Juric2008}, reducing the local halo density (relative to the local thin-disk stellar density) from $f_{h} = 0.0051$ to $f_{h} = 0.0006$, a similar value to that used by the Besan\c{c}on model \citep{Robin2003}.

A brief summary of our model, including the changes introduced in \S\ref{sec:los-scatter}, is given is \S\ref{sec:model-summary}.

Finally, we improve the convergence of our Markov Chain Monte Carlo sampling of the line-of-sight reddening profile by introducing a new type of proposal step, which we term the ``swap'' proposal. See Appendix \ref{app:swap-proposal} for details.

\subsection{3D Dust Priors}
\label{sec:dust-priors}

\begin{figure*}[htb!]
	\plotone{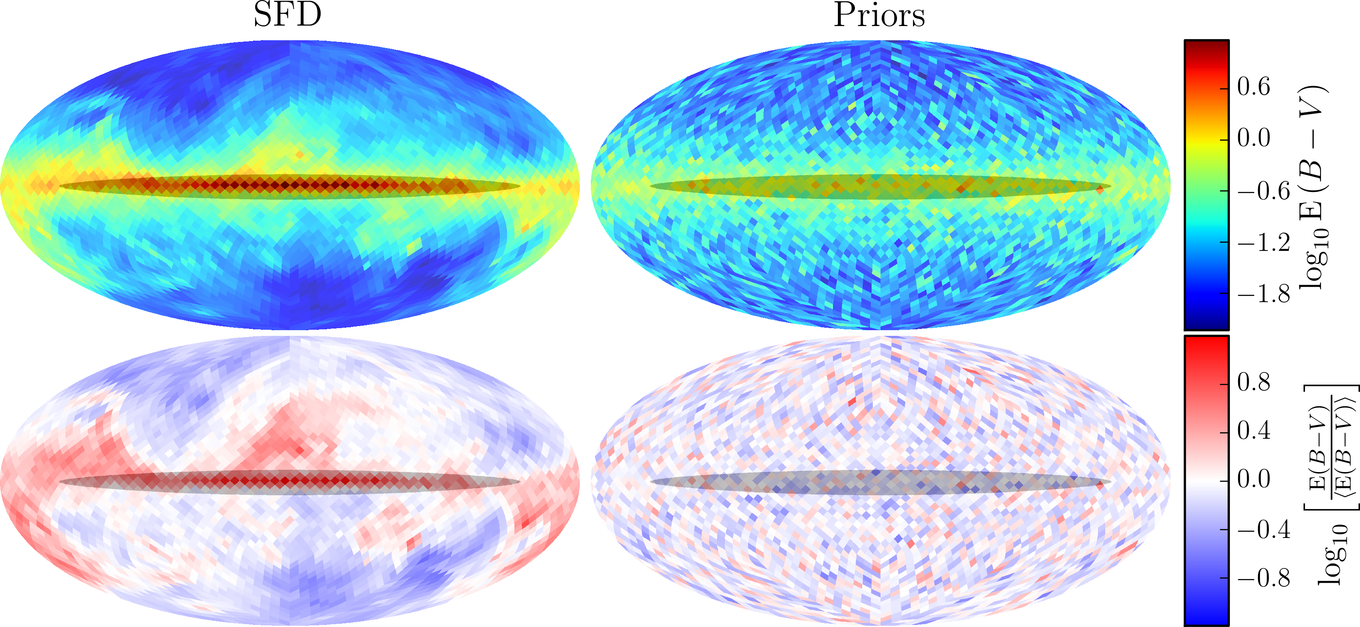}
	\caption{The SFD dust map, compared to a random draw from the priors on the 3D dust reddening distribution, used in the construction of our map. The top two panels, from left to right, show the SFD reddening and the 2D reddening map that results from a draw from the priors, both on a log scale. The bottom two panels show the same maps, after dividing out the mean projected reddening in the priors. The priors are limited so that the mean expected reddening in any given distance bin does not exceed a pre-defined amount. This is done in order to avoid inferring large amounts of reddening in the absence of data. Regions that are affected by this clipping are shaded in gray. The priors do not include spatial correlations, as can be seen by comparing the bottom two panels. \label{fig:dust-priors}}
\end{figure*}

\citet{Green2014} describes how we use photometry of stars to determine the distribution of dust in the Galaxy. We begin by separating the sky into small sightlines. In a small region of the sky across which the dust column does not vary much angularly, more distant stars should be more heavily reddened than nearby stars, as all the stars lie along the same dust column. We therefore begin by calculating a probabilistic reddening and distance estimate for each star in the sightline. We then determine the amount of dust reddening in each distance bin along the sightline, under the constraint that the line-of-sight distance vs. reddening profile has to be consistent with our distance and reddening inferences for all the stars in the sightline.

\citet{Green2014} states our map-making method generally, without choosing specific priors on the distribution of reddening in 3D. Here, we will briefly restate the formalism used in \citet{Green2014}, and then define our 3D reddening priors.

Our goal of inferring the line-of-sight dust reddening on the basis of stellar photometry begins with inferring the distance and reddening of each star based on its photometry. Label each star in the line of sight by a number $i$, and denote the probability density for an individual star to lie at distance modulus, $\mu$, and reddening, $E$, given photometry $\vec{m}$, by
\begin{align}
	p \arg{\mu_{i} , \, E_{i} \, | \, \vec{m}_{i}} \, .
    \label{eqn:indiv-stellar-pdf}
\end{align}
We precompute $p \arg{\mu_{i} , \, E_{i} \, | \, \vec{m}_{i}}$ using a kernel density estimate of Markov Chain Monte Carlo samples. This technique, similar to that used in \citet{Hogg2010}, is not guaranteed to converge to exactly the target density. It is possible, however, to substitute a mathematically correct, but somewhat more computationally expensive algorithm, \textit{pseudo-marginal} Markov Chain Monte Carlo, \citep{Beaumont2003, Andrieu2009}, in which a new noisy estimate of a marginalized likelihood term is computed at each Markov chain step. However, we expect any possible bias introduced by precomputing Eq. \eqref{eqn:indiv-stellar-pdf} once to be small due to the relatively large number of stars in each sightline. Compared to possible inaccuracies in our stellar model, for example, we expect the inexactness of our sampling method to be a small effect.

As explained in \citet{Green2014}, we place a flat prior on $E_{i}$ for each individual star when precomputing Eq. \eqref{eqn:indiv-stellar-pdf}, as the prior on reddening will come in when we combine information from all of the stars in a sightline. We now parameterize the line-of-sight reddening by a set of parameters, $\vec{\alpha}$, so that the cumulative reddening out to a distance modulus $\mu$ can be written as:
\begin{align}
	E \arg{\mu ; \vec{\alpha}} \, .
\end{align}
The probability density of $\vec{\alpha}$ is then given by a product over line integrals through the individual stellar probability density functions, following $E \arg{\mu ; \vec{\alpha}}$:
\begin{align}
	p \arg{\vec{\alpha} \, | \left\{ \vec{m} \right\}}
	&\propto p \arg{\vec{\alpha}} \, \prod_{i} \int \! \d \mu_{i} \,\, p \arg{\mu_{i} , E \arg{\mu_{i} ; \vec{\alpha}} | \, \vec{m}_i} \, . \label{eqn:line-integral-product}
\end{align}

We parameterize the reddening in each line of sight as a monotonically increasing, piecewise-linear function in distance modulus. Dividing up the line of sight into bins of equal width in distance modulus, we sample the increase in reddening in each bin, so that
\begin{align}
	\vec{\alpha} = \left\{ \Delta E_{k} \, | \, k = 1 , 2 , \ldots , N_{\mathrm{bins}} \right\} \, ,
\end{align}
where $k$ denotes the bin index, and $N_{\mathrm{bins}}$ is the number of distance bins. We place a log-normal prior on the reddening in each bin:
\begin{align}
	\Delta E_{k} &= e^{\epsilon_{k}} \, , \ \ 
	\mathrm{where} \ \epsilon_{k} \sim \mathcal{N} \arg{\bar{\epsilon}_{k} \, , \sigma_{\epsilon}} \, .
\end{align}
In each bin, we set $\bar{\epsilon}_{k}$ so that the mean reddening in each voxel matches what would be expected from the smooth disk component of \citet{Drimmel2001a}, up to a constant normalization that is the same for the entire map. In order to fully define the priors, there are therefore two global parameters that must be set:
\begin{enumerate}
    \item $\nicefrac{\mathrm{d} \EBV}{\mathrm{d} s} \big|_{s = 0}$, the local normalization of the dust reddening per unit distance, and
    \item $\sigma_{\epsilon}$, the width of the log-normal prior on dust column density in each voxel.
\end{enumerate}
We fit these two numbers so that our priors, in projection, produce two-dimensional dust maps with the same overall normalization and standard deviation as the SFD dust map. We find that $\sigma_{\epsilon} = 1.4$ and a local reddening per unit distance of $\nicefrac{\mathrm{d} \EBV}{\mathrm{d} s} = 0.2 \, \mathrm{mag \, kpc^{-1}}$ roughly matches the mean and variance of the SFD dust map over large angular scales.

Fig. \ref{fig:dust-priors} shows a random realization of our 3D dust priors, next to the SFD dust map. In each bin, we limit $\bar{\epsilon}$ to the range $-12 \leq \bar{\epsilon} \leq -4$. On the lower end, the limit helps our fit to converge by preventing the prior from becoming too stringent. On the upper end, the limit prevents our fit from inferring large amounts of dust in the absence of data. Regions where $\bar{\epsilon}$ has been limited to $-4$ are shaded in Fig. \ref{fig:dust-priors}. Our priors do not impose spatial correlations across lines of sight, and thus in the absence of data, produce cloud-free maps. The detailed cloud structures that emerge from our 3D dust modeling therefore derive entirely from the data, rather than from the priors.

It is worth noting that our priors diverge from what one should expect for the real distribution of dust in a number of ways. In order to render the problem of fitting the 3D distribution of dust tractable, we fit the dust column along each sightline separately. We know, however, that in real life, dust density has spatial correlations. Moreover, there is nothing special about the Sun's place in the Galaxy, but our dust map voxelizes the sky into pencil beams centered on the Solar System. This of course makes practical sense, since angles are much easier to measure than distances in astronomy, but it is again an unreal feature of our voxelization. A more realistic model would treat the dust density as a continuous field, or voxelize the Galaxy in a way that treats the angular and radial directions equally, and would impose correlations between dust density in nearby points in space \citep[see, e.g.,][]{Lallement2013,Sale2014a}. This entails significantly more algorithmic and computational complexity than the method used here, and we defer such work to the future. Within the constraints of our present setup -- independent sightlines with pencil-beam-like voxels -- our priors attempt to reasonably trace the properties of the Galaxy, including the mean dust density in each voxel and the overall variance in dust column across the sky.

\subsection{Scatter in Line-of-Sight Reddening Profile}
\label{sec:los-scatter}

\begin{figure}[!htpb]
	\plotone{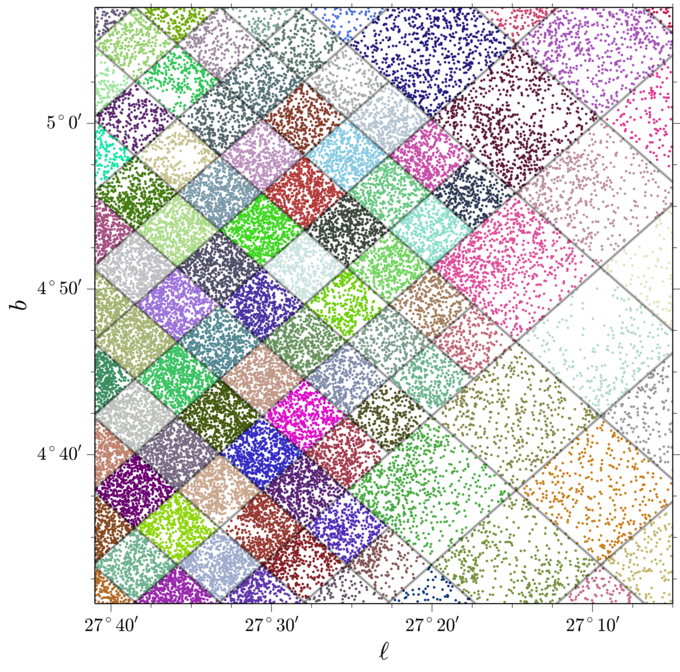}
	\caption{Assigning stars to different pixels, in which the line-of-sight dust profile will be fit independently. Here, each dot represents a point-source detection, and the dots are colored by the HEALPix pixel they are assigned to. The pixel scale varies in order to keep the number of stars per pixel roughly constant. Note that some pixels show significant extinction-induced variations in stellar density. \label{fig:pixelization}}
\end{figure}

A basic assumption of our model, as described in \citet{Green2014}, is that within a given pixel, the dust density varies only with distance, and not with angle. If the pixels are sufficiently small, this is a good assumption. We are limited, however, in how small we can make each pixel by the need to include enough stars in each pixel to probe the dust density at a range of distances. Increasing the angular resolution of the map decreases the number of stars in each pixel, effectively decreasing the distance resolution of the map. We have found that we obtain best results for pixels containing a few hundred stars, and we vary the resolution of our pixels across the sky to obtain approximately the same number of stars in each pixel (see Fig. \ref{fig:pixelization}).

Note that varying the pixel size across the sky in this way technically violates the principles of Bayesian inference. From a forward-modeling point of view, the distribution of dust influences the number of stars that are observed in any given region of the sky. \citet{Sale2015}, for example, discusses how a catalog of stars observed in a survey can be described as a Poisson point process whose rate is determined by the distribution of stars in the Galaxy, the three-dimensional distribution of dust, and the survey selection function. Yet we are using the number of stars observed in each part of the sky to determine how to pixelize the sky. Our pixelization is therefore set, in part, by an observable (the density of stars across the sky) that is a consequence of the model. Nevertheless, as we are treating each pixel independently, and we would like to keep pixels as small as possible without reducing the number of stars per pixel below a few hundred, this rules violation is difficult to avoid. We expect the impact of this violation to be small in most regions of the sky. In regions with large sub-pixel variation in dust reddening, however, our map may be biased towards lower reddenings, as we preferentially detect stars in regions of the pixel with lower reddening.

The typical resolution of the map is $6.8^{\prime} \times 6.8^{\prime}$, corresponding to an $\mathtt{nside} = 512$ HEALPix pixelization \citep{Gorski2005}. At this resolution, there can still be significant power in the dust density spectrum below the pixel scale. This can pose problems for our method, especially in the vicinity of dense clouds and filamentary structures, where the sub-pixel angular variation is largest. In order to deal with sub-pixel angular variation in the dust density, we relax our assumption that all stars lie along the same dust column by allowing each star to deviate from the local ``average'' dust column by a small amount. The reddening of star $i$ is parameterized as
\begin{align}
	E_{i} &= \left( 1 + \delta_{i} \right) E \arg{\mu_{i} ; \vec{\alpha}} \, ,
    \label{eqn:E-delta-relation}
\end{align}
where $\delta_{i}$ is the fractional offset of the star from the local dust column, $E \arg{\mu_{i} ; \vec{\alpha}}$.

In effect, our model is therefore that within each HEALPix pixel, the reddening is a white noise process, with a mean that increases piecewise linearly with distance. Each star samples this white noise process at a particular distance and angular position within the pixel. The parameter $\delta_{i}$ is then understood as the fractional residual (from the mean reddening in the pixel at the given distance) of the reddening column at the angular location and distance of star $i$.

We put a Gaussian prior on $\delta_{i}$, with zero mean and standard deviation dependent to the scale of the pixel (allowing more variation in larger pixels) and the local dust column (allowing greater fractional variation in regions of greater reddening). In detail,
\begin{align}
	p \arg{\delta_{i} \, | \, \mu_{i}, \, \vec{\alpha}}
	&= \mathcal{N} \arg{ \delta_{i} \, | \, 0, \, \sigma_{\delta} } \, ,
    \label{eqn:scatter-prior}
\end{align}
with
\begin{align}
	\sigma_{\delta} = a E \arg{\mu_{i} ; \vec{\alpha}} + b \, .
\end{align}
Here, $a$ and $b$ are parameters that we set in order to match the variation we see at the given pixel angular scale in the Planck radiance-based two-dimensional dust map. We compute the RMS scatter within HEALPix pixels of difference scales, finding that the scatter is well described by setting the coefficients $a$ and $b$ to
\begin{align}
    \log_{10} a &= 0.88 \log_{10} \left( \frac{\varphi}{1^{\prime}} \right) - 2.96 \, , \\
    \log_{10} b &= 0.58 \log_{10} \left( \frac{\varphi}{1^{\prime}} \right) - 1.88 \, ,
\end{align}
where $\varphi$ is the angular pixel scale, defined as the square-root of the pixel solid angle.

We have found through trial and error that it is preferable to impose a minimum scatter of $\sim 10\%$ on the in-pixel dust column. Additionally, if we allow $\delta_{i}$ to approach unity, we clearly risk the possibility of scattering a star to negative dust column. We therefore never allow $\sigma_{\delta} > 0.25$.

Although the complication introduced in this section adds an additional parameter, $\delta_{i}$, for each star, it can be achieved with minimal additional computational resources. We are introducing a Gaussian scatter in the reddening of each star from the ``average'' reddening in the pixel, and an appropriate Gaussian smoothing of the individual stellar probability density surfaces, $p \arg{ \mu_{i} , \, E_{i} \, | \, \vec{m}_{i}}$, achieves this effect. Appendix \ref{app:los-scatter} details how the individual stellar probability surfaces are smoothed.

\subsection{Summary of Model}
\label{sec:model-summary}

In summary, our model of each sightline contains the following elements:
\begin{itemize}
    \item The increase in the ``average'' reddening in each distance bin: $\vec{\alpha} = \left\{ \Delta E_{k} \, | \, \, k = 1, \ldots , n_{\mathrm{bins}} \right\}$.
    \item The distance modulus, $\mu_{i}$, stellar type, $\Theta_{i}$ and fractional offset, $\delta_{i}$ from the ``average'' reddening of each star $i$. 
\end{itemize}
The reddening of star $i$ is therefore determined both by $\vec{\alpha}$ and $\delta_{i}$. Together with the distance and type of the star, one can obtain model apparent magnitudes for the star, and thus a likelihood:
\begin{align}
    p \arg{\vec{m}_{i} \, | \, \mu_{i} , \vec{\Theta}_{i} , \delta_{i} , \vec{\alpha}} \, .
\end{align}
We also have per-star priors on distance and stellar type, given by a smooth model of the distribution of stars throughout the Galaxy, and a prior on the offset of each star from the local reddening column:
\begin{align}
    p \arg{\mu_{i} , \vec{\Theta}_{i} , \delta_{i} \, | \, \vec{\alpha}}
    &= p \arg{\mu_{i} , \vec{\Theta}_{i}} p \arg{\delta_{i} \, | \, \mu_{i} , \vec{\alpha}} \, .
\end{align}
We finally have a log-normal prior on the increase in reddening in each distance bin along each sightline, $p \arg{\vec{\alpha}}$, whose mean is chosen to match a smooth model of the distribution of dust throughout the Galaxy.

The posterior on the line-of-sight reddening along one sightline is then given by
\begin{align}
    p \arg{\vec{\alpha} \, | \left\{ \vec{m} \right\}}
    &= p \arg{\vec{\alpha}} \prod_{i=1}^{n_{\mathrm{stars}}}
    p \arg{\vec{m}_{i} \, | \, \mu_{i} , \vec{\Theta}_{i} , \delta_{i} , \vec{\alpha}}
    \notag \\ &\hspace{2cm} \times
    p \arg{\mu_{i} , \vec{\Theta}_{i}} p \arg{\delta_{i} \, | \, \mu_{i} , \vec{\alpha}} \, .
\end{align}
We pre-compute the likelihood and prior terms for the individual stars, and then sample in $\vec{\alpha}$. The details of how this is done are given in \citet{Green2014}, and in Appendix \ref{app:los-scatter} of this paper.

\subsection{Improving Stellar Inferences using the 3D Map}
\label{sec:reweighted-stellar-samples}

In order to create the 3D dust map, we first probabilistically infer the distance and reddening to each star individually, with no information about the 3D structure of the dust. After creating the 3D dust map, however, we have a very strong constraint on how reddening should increase with distance, and this should impact our inferences about individual stellar distances and reddenings. In order to improve our stellar parameter inferences, we should replace our initial assumption about stellar reddening (a flat prior) with a new one that favors stellar reddenings close to the measured distance-reddening relation along the sightline. This can be done, in practice, by reweighting the probability density functions we initially calculated for each star. In this section, we therefore define a reweighting of the Markov Chain samples for the individual stars which takes into account the line-of-sight reddening.

\begin{figure}[htb!]
	\plotone{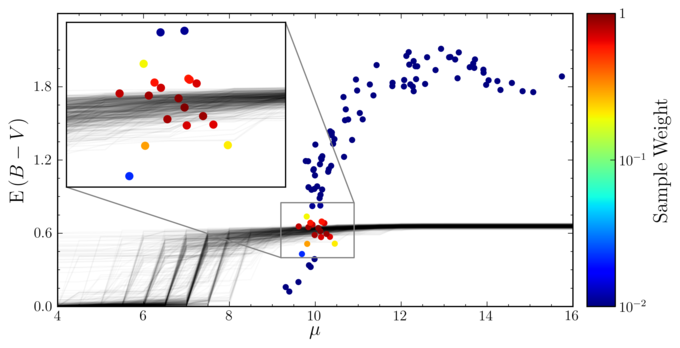}
	\caption{Reweighting Markov Chain samples for an individual star, based on the line-of-sight reddening profile. Each black line is a reddening profile drawn from the posterior on line-of-sight reddening. The dots are posterior samples of parameters (including distance and reddening) for one star viewed in isolation, i.e., not conditioned on the line-of-sight reddening. The samples are reweighted in order to condition them on the line-of-sight reddening, assigning greater weight to samples that are consistent with the reddening profile. Including information about the line-of-sight reddening can significantly reduce the uncertainties in stellar distance, reddening and type. \label{fig:reweighted-stellar-samples}}
\end{figure}

Let us first consider the case in which we fix the line-of-sight reddening profile. In the formalism used here, that means that we fix the parameters $\vec{\alpha}$, expressing the line-of-sight reddening profile as $E \arg{\mu ; \vec{\alpha}}$. The reddening of an individual star is then determined by the distance modulus of the star, $\mu$, as well as the fractional deviation, $\delta$, of the stellar reddening from the local dust column. As before, there are also parameters representing the stellar type, $\vec{\Theta}$. In this formulation, the posterior density of the stellar parameters, given its photometry $\vec{m}$ and the line-of-sight reddening profile, is determined by
\begin{align}
    p \arg{\mu , \vec{\Theta} , \delta \, | \, \vec{m} , \, \vec{\alpha}}
    &\propto p \arg{\vec{m} \, | \, \mu , \vec{\Theta} , \delta , \vec{\alpha}} p \arg{\mu , \vec{\Theta}} p \arg{\delta \, | \, \mu , \vec{\alpha}} \, .
\end{align}
We already have Markov chain samples in distance, reddening and stellar type for each star, for a model which does not take the line-of-sight reddening into account. We would like to apply weights to these samples, so that they correspond to the model sketched out directly above. As shown in detail in Appendix \ref{app:rw-samples-detailed}, the correct reweighting of the stellar Markov chain samples, assuming a particular line-of-sight reddening profile $\vec{\alpha}$, is
\begin{align}
    w_{k} \propto \frac{ \mathcal{N} \arg{\delta_{k} \, | \, 0 , \sigma_{\delta}} }{ E \arg{\mu_{k} ; \vec{\alpha}} } \, ,
\end{align}
where $k$ indexes the sample, and $\delta_{k}$ is the fractional offset of the sample reddening, $E_{k}$, from the line-of-sight reddening, $E \arg{\mu_{k}; \vec{\alpha}}$, at distance $\mu_{k}$. The numerator in the above weight corresponds to a prior on the offset of the stellar reddening from the local dust column, while the denominator comes from the Jacobian transformation from reddening to fractional offset from the local reddening.

Reweighting each of the stellar parameter samples by the above factor, and normalizing the sum of the weights to unity, we obtain an inference for the stellar parameters, conditioned on a particular line-of-sight reddening profile, $E \arg{\mu ; \vec{\alpha}}$. We marginalize over the line-of-sight reddening profile by repeating this procedure for each sampled value of $\vec{\alpha}$, summing the weight applied to each stellar sample.

It might be objected that since we infer $\vec{\alpha}$ using the photometry of all the stars in the given pixel, the samples we drew for $\vec{\alpha}$ in our initial processing are already dependent on the photometry for the star whose samples we are reweighting. Put more formally, the prior we use here when marginalizing over line-of-sight reddenings is the inference we obtained earlier, $p \arg{\vec{\alpha} \, | \left\{ \vec{m} \right\}}$, which is conditional on all the photometry in the pixel, $\left\{ \vec{m} \right\}$. We are using the photometry of a given star to infer the line-of-sight reddening, and then again to infer the stellar parameters, conditional on that line-of-sight reddening.

This would be a problem if the photometry of any given star significantly affected the line-of-sight reddening inference. However, if photometry from a large number of stars informs the line-of-sight reddening, then no one star should have a significant impact on the inferred line-of-sight reddening profile. In a more formally correct formulation of the problem, we would first infer the line-of-sight reddening using all the stars in the pixel except the star whose parameters we wish to infer, and we would then infer the parameters for that star, conditional on the inferred line-of-sight reddening profile. As each pixel contains hundreds of stars, however, we expect our procedure to approximate this formally correct procedure closely.

Fig. \ref{fig:reweighted-stellar-samples} shows the result of reweighting the samples for one star. Depending on the line-of-sight reddening profile and the distribution of the unweighted stellar parameter samples, individual stellar inferences can be tightened dramatically by taking the line-of-sight reddening into account. Our knowledge of the parameters describing one star is therefore dependent not only on its photometry, but also on the photometry of its neighbors on the sky. Because neighboring stars lie along the same dust column, inferences for nearby stars are coupled through the requirement that the dust column increase with distance.

\section{Data}
\label{sec:data}

\subsection{Pan-STARRS 1}

Pan-STARRS 1 is a 1.8-meter optical and near-infrared telescope located on Mount Haleakala, Hawaii \citep{PS1_system, PS1_optics}. The telescope is equipped with the GigaPixel Camera 1 (GPC1), consisting of an array of 60 CCD detectors, each 4800 pixels on a side \citep{PS1_GPCa, PS1_GPCb}. From May 2010 to April 2014, the majority of the observing time was dedicated to a multi-epoch $3 \pi$ steradian survey of the sky north of $\delta = -30^{\circ}$ (Chambers in prep.). The $3 \pi$ survey observes in five passbands $g_{\mathrm{P1}}$, $r_{\mathrm{P1}}$, $i_{\mathrm{P1}}$, $z_{\mathrm{P1}}$, and $y_{\mathrm{P1}}$, similar to the Sloan Digital Sky Survey \citep[SDSS;][]{York2000}, with the most significant difference being the replacement of the Sloan $u$ band with a near-infrared band, $y_{\mathrm{P1}}$. The PS1 filter set spans 400--1000~nm \citep{PS_lasercal}. The images are processed by the Pan-STARRS 1 Image Processing Pipeline (IPP) \citep{PS1_IPP}, which performs automatic astrometry \citep{PS1_astrometry} and photometry \citep{PS1_photometry}. The data is photometrically calibrated to better than 1\% accuracy \citep{Schlafly2012, JTphoto}. The $3 \pi$ survey reaches typical single-exposure depths of 22~mags (AB) in $g_{\mathrm{P1}}$, 21.5~mags in $r_{\mathrm{P1}}$ and $i_{\mathrm{P1}}$, 20.8~mags in $z_{\mathrm{P1}}$, and 20~mags in $y_{\mathrm{P1}}$. The resulting homogeneous optical and near-infrared coverage of three quarters of the sky makes the Pan-STARRS1 data ideal for studies of the distribution of the Galaxy's dust.

\subsection{Two Micron All Sky Survey}

The Two Micron All Sky Survey (2MASS) is a uniform all-sky survey in three near-infrared bandpasses, $J$, $H$ and $K_{s}$ \citep{Skrutskie2006}. The survey derives its name from the wavelength range covered by the longest-wavelength band, $K_{s}$, which lies in the longest-wavelength atmospheric window not severely affected by background thermal emission \citep{Skrutskie2006}. The survey was conducted from two 1.3-meter telescopes, located at Mount Hopkins, Arizona and Cerro Tololo, Chile, in order to provide coverage for both the northern and southern skies, respectively. The focal plane of each telescope was equipped with three $256 \times 256$ pixel arrays, with a pixel scale of $2^{\prime \prime} \times 2^{\prime \prime}$. Each field on the sky was covered six times, with dual 51-millisecond and 1.3-second exposures, achieving a $10 \sigma$ point-source depth of approximately 15.8, 15.1 and 14.3~mag (Vega) in $J$, $H$ and $K_{s}$, respectively. Calibration of the survey is considered accurate at the 0.02~mag level, with photometric uncertainties for bright sources below 0.03~mag \citep{Skrutskie2006}.

The addition of 2MASS data requires an expansion of the stellar model described in \citet{Green2014} to cover the 2MASS $J$, $H$ and $K_{s}$ passbands, as well as an estimate of the survey selection function for 2MASS. We address these additions to our model in Appendices \ref{app:PS1-2MASS-models} and \ref{app:2MASS-selection}.

\subsection{Source Selection}

We match each PS1 source to the nearest source in the 2MASS point-source catalog, rejecting stars at greater than $2^{\prime \prime}$ separation. We require detection in at least four passbands, two of which must be PS1 passbands. In order to reject extended sources, we require that $m_{\mathrm{psf}} - m_{\mathrm{aperture}} < 0.1 \, \mathrm{mag}$ in at least two PS1 passbands. We additionally reject sources flagged as extended sources in 2MASS.

Individual PS1 passbands are rejected if they have photometric uncertainty greater than $0.2 \, \mathrm{mag}$, or if the sources are beyond or close to the saturation limit for PS1 (here considered $14.5$, $14.5$, $14.5$, $14$ and $13 \, \mathrm{mag}$ in $grizy_{\mathrm{P1}}$, respectively). In 2MASS passbands, we make the recommended ``high-reliability catalog'' selection cuts,\footnote{See the 2MASS All-Sky Data Release Explanatory Supplement: \url{http://www.ipac.caltech.edu/2mass/releases/allsky/doc/sec1_6b.html\#composite}} in addition to the following requirements:

\begin{itemize}
    \item $\mathtt{contamination / confusion \ flag} = 0$,
    \item $\mathtt{galaxy \ contamination \ flag} = 0$.
\end{itemize}

Our final catalog contains 798,611,689 sources, of which 32\% are detected in four passbands, 49\% in five passbands, and 19\% in six or more passbands.

\subsection{Pixelization}

\begin{deluxetable}{c c c c}[htb!]
	\tablewidth{0pt}
	\tablecolumns{2}
	\tabletypesize{\scriptsize}
	\tablecaption{Pixelization \label{tab:thresholds}}
	\tablehead{
	           \colhead{} \vspace{-0.15cm} &
	           \colhead{Max.} &
	           \colhead{Solid Angle} &
	           \colhead{\# of}
	           \\
	           \colhead{\texttt{nside}} \vspace{-0.15cm} &
	           \colhead{} &
	           \colhead{} &
	           \colhead{}
	           \\
	           \colhead{} &
	           \colhead{Stars / Pixel} &
	           \colhead{at Resolution ($\mathrm{deg}^{2}$)} &
	           \colhead{Pixels}
	           }
	\startdata
		64    &  200   &  77     &  91       \\
		128   &  250   &  90     &  430      \\
		256   &  300   &  11980  &  228373   \\
		512   &  800   &  16071  &  1225471  \\
		1024  &  1200  &  2957   &  901971   \\
		2048  &  ---   &  66     &  80956    \\
        \colrule \\[-5pt]
		total &  ---   &  31240  &  2437292
	\enddata
\end{deluxetable}

We divide the sky into HEALPix pixels \citep{Gorski2005}, adjusting the pixel scale in order to keep the number of stars per pixel roughly constant (see Fig. \ref{fig:pixelization}). Our procedure is to begin with $\mathtt{nside} = 64$ pixels, and then subdivide each pixel recursively, as long as the number of stars exceeds some threshold, dependent on the pixel scale. We use thresholds given in Table \ref{tab:thresholds}, chosen to allow us to reach a resolution of $\mathtt{nside} = 512$ with a relatively small number of stars, but to avoid going to higher resolutions unless the stellar density is much higher. We reject pixels with fewer than ten stars. Such pixels comprise a negligible fraction of the sky. In all, we assign just under 800 million stars, covering just over three quarters of the sky, to 2.4 million pixels, with an average of 327 stars per pixel.

\section{Results}
\label{sec:results}

\begin{figure*}[htb!]
	\plotone{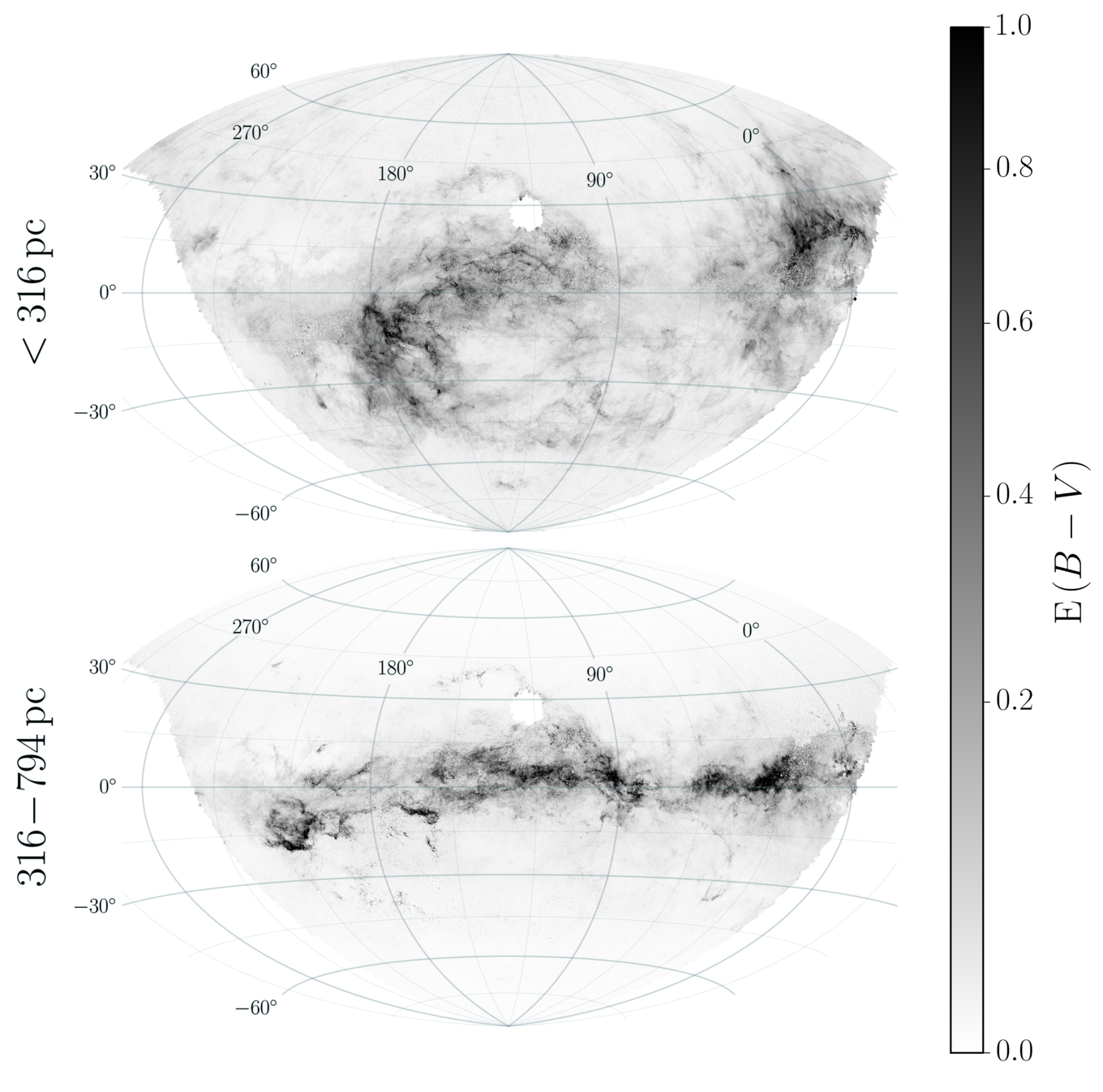}
	\caption{Median differential reddening in two Solar-centric distance ranges. The distance breaks are chosen to coincide with distance moduli $\mu = 7.5$ and 9.5, which line up with edges of distance bins in our map. We adopt a square-root stretch, in order to capture both low- and high-reddening features. The hole at $\ell \approx 120^{\circ}$, $b \approx 30^{\circ}$ corresponds to declinations above $\sim \! 84^{\circ}$, which had not yet been fully processed at the time we created our 3D map. \label{fig:dust-slices-near}}
\end{figure*}

\begin{figure*}[htb!]
	\plotone{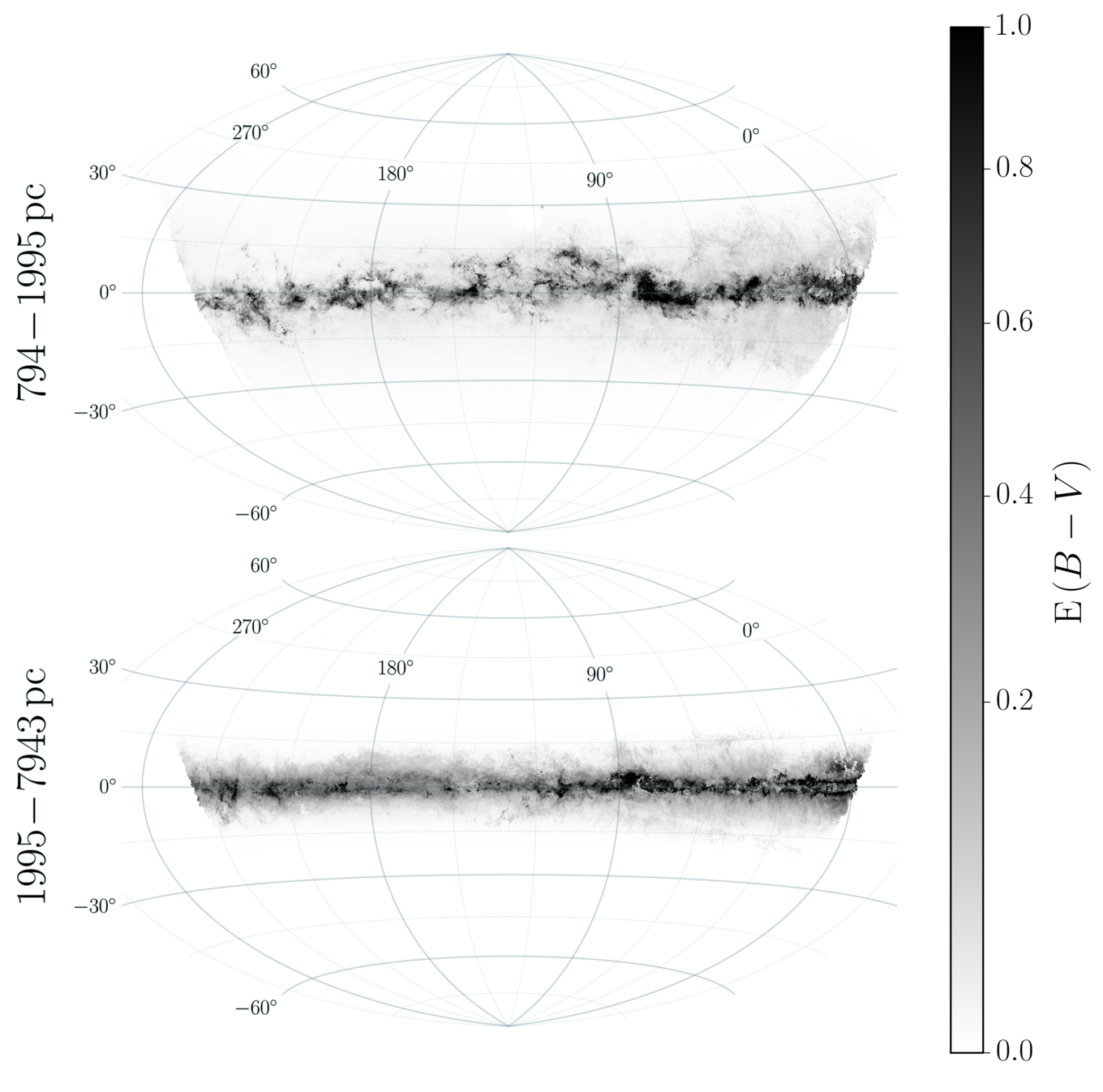}
	\caption{As in Fig. \ref{fig:dust-slices-near}, but with breaks at distance moduli $\mu = 11.5$ and 14.5. \label{fig:dust-slices-far}}
\end{figure*}

\begin{figure*}[htb!]
	\plotone{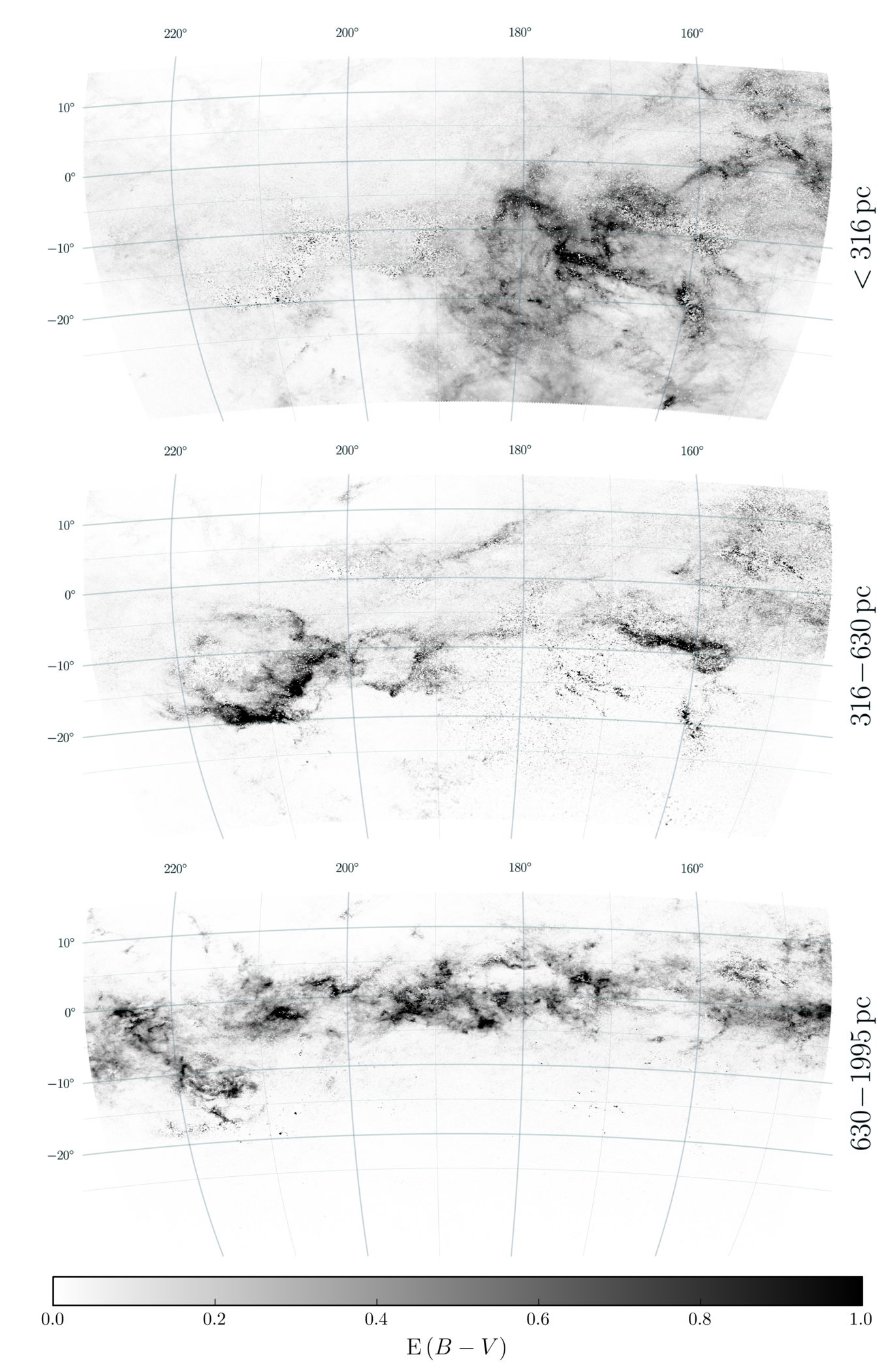}
	\caption{A closer view of the dust in the anticentral region in three distance bins. As in Figs. \ref{fig:dust-slices-near} and \ref{fig:dust-slices-far}, we plot the median differential reddening. The Taurus-Perseus-Auriga complex is visible in the right half of the nearest distance bin. In the second distance bin, the Orion complex is visible on the left, while the California cloud is visible on the right. Note the ring-like shape of the Orion complex, which is only revealed by 3D mapping when confusion from background dust is removed. See \citet{Schlafly2014Orion} for discussion of this feature. Monoceros R2 appears beyond Orion, in the third distance bin, flanked by the plane of the Galaxy at yet greater distance. \label{fig:dust-slices-detail-anticenter}}
\end{figure*}

\begin{figure*}[htb!]
	\plotone{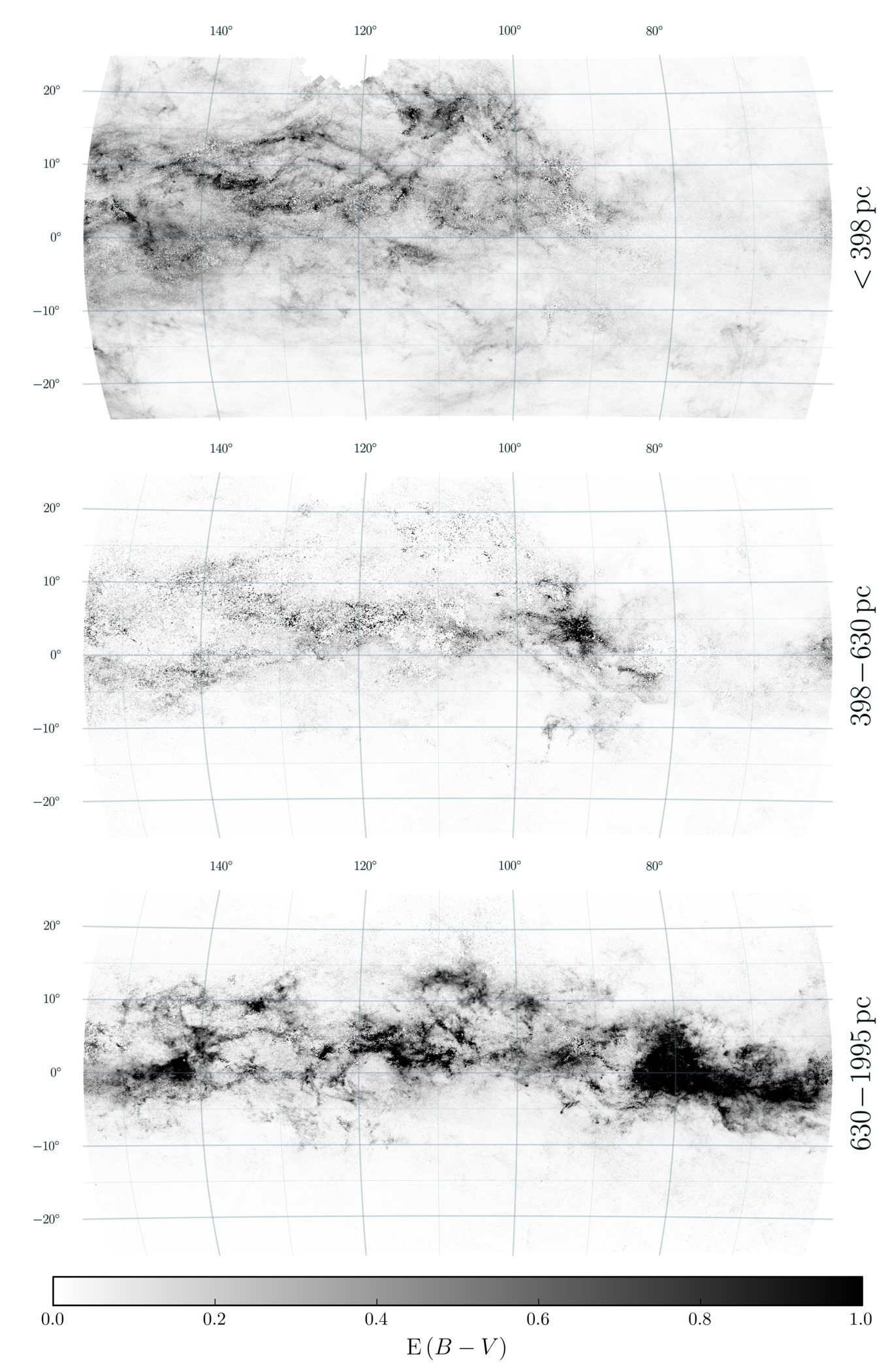}
	\caption{A closer view of the dust in the direction of Cepheus and Polaris flares and the eastern portion of the Great Rift, including Cygnus X. The Cepheus flare ($95^{\circ} \lesssim \ell \lesssim 110^{\circ}$, $b \approx 15^{\circ}$) separates into two components in distance, visible in the first and third panels. The dust associated with the Cygnus X region ($\ell \approx 80^{\circ}$, $b \approx 0^{\circ}$) appears in the third panel. \label{fig:dust-slices-detail-PolarisCepheus}}
\end{figure*}

Applying the method described in \citet{Green2014}, with the modifications described above, to Pan-STARRS 1 and 2MASS stellar photometry, we produce a three-dimensional map of dust reddening, covering the three-quarters of the sky north of $\delta = -30^{\circ}$.

\subsection{Distance Slices of Map}

In Figs. \ref{fig:dust-slices-near} and \ref{fig:dust-slices-far}, we present the differential reddening in four spherical shells, centered on the Sun. Each panel shows the median dust reddening in a different range of Solar-centric distances. Due to perspective, dust at high Galactic latitudes resides nearby, as Galactic dust lies in a thin disk. We recover the wealth of structure seen in the ISM across a wide range of scales, from thin filaments to large cloud complexes. Large, coherent cloud complexes appear at consistent distances.

Fig. \ref{fig:dust-slices-detail-anticenter} gives a closer view of the anticentral region ($\ell \sim 180^{\circ}$). Different features appear clearly in each rendered distance slice. The Perseus, Taurus and Auriga cloud complexes dominate the anticentral region in the closest distance slice, while the Orion molecular cloud complex ($\ell \sim 210^{\circ}$, $b \sim -15^{\circ}$) and the California nebula ($\ell \sim 160^{\circ}$, $b \sim -8^{\circ}$) appear very strikingly in the second distance slice. Of particular interest is the ring-like structure that Orion A and B appear to be embedded within. This ring-like structure is only apparent when the background dust is removed. In particular, the northeast portion of the ring is confused with the plane of the Galaxy in projection. \citet{Schlafly2014Orion} discusses the ``Orion ring,'' including possible formation scenarios for the ring, in greater depth.

Fig. \ref{fig:dust-slices-detail-PolarisCepheus} shows the Galactic plane from $\ell = 60^{\circ}$ to $155^{\circ}$ in more detail. The Cepheus flare, which lies at the center of the frame, at $95^{\circ} \lesssim \ell \lesssim 110^{\circ}$, $b \approx 15^{\circ}$, separates into two components at different distances, visible in the first and third panels. Using a modified version of the method used in this paper along individual sightlines, \citet{Schlafly2014clouds} places the two components of the Cepheus flare at distances of $360 \pm 35 \pc$ and $900 \pm 90 \pc$. The dust associated with Cygnus X ($\ell \approx 80^{\circ}$, $b \approx 0^{\circ}$) appears in the third panel, along with a wealth of fine structure along the Galactic plane.

We note that each pixel is fit independently, and our only prior assumption about the spatial structure of the dust is that if forms a diffuse disk, as shown in Fig. \ref{fig:dust-priors}. The detailed spatial structure in the interstellar medium that our analysis reveals indicates that the PS1 and 2MASS photometry dominates over our priors out to a distance of several kiloparsecs and reddening of $\EBV \approx 1.5 \, \mathrm{mag}$. With the assumption of spatial correlations between neighboring pixels, we expect that one would be able to significantly reduce the uncertainty in the map, and achieve better distance resolution \citep[see, e.g.,][]{Sale2014a}.

\subsection{Maximum \& Minimum Reliable Distances in Map}
\label{sec:reliable-dists}

Our 3D dust map is based on measurements of stellar distances and reddenings. Beyond the most distant stars, we have no sensitivity to dust, and in front of the nearest stars, we have no information about the distance to the dust. Therefore, we estimate the minimum and maximum distance to which our map is reliable by locating the nearest and farthest stars in each pixel. Outside of this distance range, our line-of-sight reddening inferences are dominated by our priors. Using the improved stellar parameter inferences (described in \S\ref{sec:reweighted-stellar-samples}), we define the minimum reliable distance in each line of sight as the distance out to which there are $N_{\mathrm{closer}}$ observed main-sequence stars, and the maximum reliable distance as the distance beyond which there are $N_{\mathrm{farther}}$ observed main-sequence stars.

For this calculation, we count each Markov Chain sample in stellar distance as a fraction of a star. We exclude stars that fail to converge, for which the model is a very poor match to the data \citep[as determined by the Bayesian evidence for the star; see][]{Green2014}, or which are inconsistent with the inferred line-of-sight reddening profile. We consider a star inconsistent with the line-of-sight reddening inference if none of the 100 stored Markov Chain samples of the stellar distance and reddening is within a fractional distance $\sigma_{\delta}$ (the modeled intra-pixel scatter in the dust column; see \S\ref{sec:reweighted-stellar-samples}) of the line-of-sight reddening profile. Such objects are likely not well fit by any of our stellar templates, or alternatively signal that there is more variation in the dust column at fine angular scales than we allow.

In determining the minimum and maximum reliable distances in the 3D dust map, we use only main-sequence stars. This is because we consider our inferences for giants to be less reliable than our inferences for dwarfs. The colors and luminosities of giants depend more strongly on metallicity and age, the latter of which we do not model.

\begin{figure*}[htb!]
	\plotone{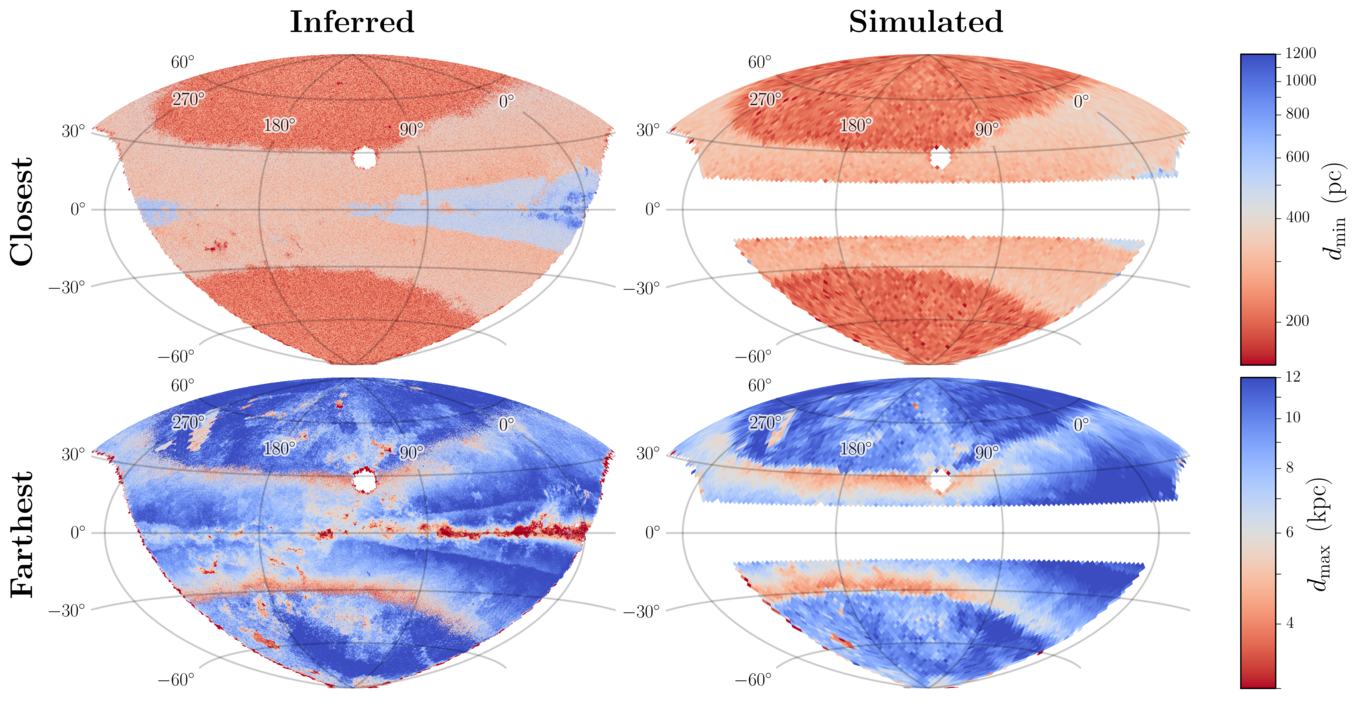}
	\caption{The minimum and maximum reliable distances in the 3D dust map. The left two panels are based on the inferred distribution of stars along each line of sight, while the right two panels are based on simulated stellar catalogs generated from our Galactic priors. We consider the map reliable if there are at least two stars closer than the given distance, and at least ten stars beyond the distance. Pixel size, survey depth and completeness, stellar density throughout the Galaxy and the presence of dust all affect the distribution of observed stars along the line of sight, and therefore the distance range over which the map is reliable. Sharp transitions in reliable distance occur at pixel size boundaries, where the number of stars per pixel changes discontinuously, as well as in patches of the sky with missing PS1 passbands. \label{fig:map-depth}}
\end{figure*}

In the left two panels of Fig. \ref{fig:map-depth}, we show the results obtained by requiring $N_{\mathrm{closer}} = 2$ and $N_{\mathrm{farther}} = 10$. The results are qualitatively similar for other choices of these parameters, as long as they are small compared to the typical number of stars in a pixel. The closest reliable distance is set almost entirely by the angular pixel scale. The nearby density of stars is, to a very rough estimation, uniform, meaning that the distance to the closest star in a pixel is a function primarily of the solid angle covered by the pixel. Accordingly, the top left panel of Fig. \ref{fig:map-depth} is essentially a map of pixel solid angle, with boundaries in distance following pixel scale boundaries.

The farthest reliable distance of the 3D dust map is strongly influenced not only by the pixel scale, but also by the distribution of stars throughout the Galaxy, the 3D distribution of dust and the survey depth. The survey depth plays a larger role in the far limit because the most distant observed stars lie preferentially near the limiting survey magnitude. The closest stars, by contrast, are distributed more evenly in apparent magnitude. Accordingly, the high-latitude patches with particularly shallow maximum reliable distances (e.g., the shallow patch near $\ell = 205^{\circ}$, $b = -55^{\circ}$) correspond to areas which are missing one or more PS1 bands, or which have fewer observation epochs in the PS1 $3\pi$ survey, and therefore worse coverage.

As a check on these estimated nearest and farthest reliable distance maps, we construct equivalent maps using simulated stellar catalogs. For each pixel in our map, we use our Galactic priors and the estimated limiting magnitudes for $grizy_{\mathrm{P1}}$ and 2MASS $JHK_{s}$ to generate an equal number of stars as actually observed. We then determine the distance in front of which there are $N_{\mathrm{closer}}$ and beyond which there are $N_{\mathrm{farther}}$ main-sequence stars. In order to remove the complicating factor of the line-of-sight dust inference, we restrict this test to $\left| b \right| > 15^{\circ}$ and to pixels for which $\EBV_{\tau_{353}} < 0.05 \, \mathrm{mag}$, and assume for our simulated catalogs that there is zero dust extinction. In these comparisons, we find that a halo number density close to the value given in \citet{Juric2008} is required to reproduce inferred map depths of Fig. \ref{fig:map-depth}. The results for a halo strength of $n_{h} = 0.004$, binned down to $\mathtt{nside} = 64$, are shown in the right two panels of Fig. \ref{fig:map-depth}.

These results indicate that while we dialed down the halo strength in our priors, the data, as reflected in our photometric stellar inferences, nonetheless prefers a stronger halo. In future work, we will investigate the implications of our stellar and line-of-sight dust inferences for a global Galactic structure model.

\subsection{Stellar Inferences}

As described in \S\ref{sec:reweighted-stellar-samples}, we reweight the naively inferred parameters for each star in order to take the line-of-sight reddening profile into account. In order to demonstrate the improvement in stellar inferences after this re-weigthing procedure, we compare our reddening inferences to a set of independently determined stellar reddening standards, as in \citet{Green2014}. For this, we use the set of stellar reddening standards developed by \citet{Schlafly2011}. The Sloan Extension for Galactic Understanding and Exploration \citep[SEGUE;][]{Yanny2009}, part of SDSS-II, used moderate-resolution spectra to classify 240,000 stars. Empirically adjusted SDSS colors based on model atmospheres can then be compared with observed SDSS photometry to obtain reliable reddening estimates. We select the same sample of 200,000 SEGUE target stars as \citet{Schlafly2011}, which excludes white dwarfs and M dwarfs (the former because they are not contained in our model, and the latter because of their unreliable spectral classification).

\begin{figure}[htb!]
	\plotone{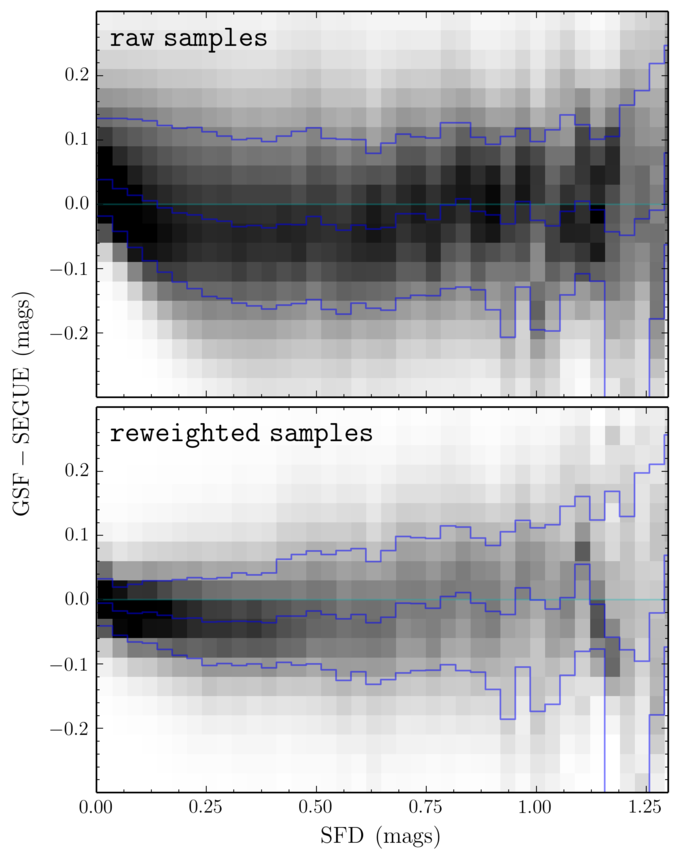}
	\caption{Histograms of the residuals of our PS1+2MASS-based stellar reddenings (referred to as ``GSF'' here), versus reddening estimates obtained by comparing SEGUE spectral classifications with SDSS photometry. The histograms are spread out along the $x$-axis by the local SFD reddening. The blue lines trace the 16th, 50th and 84th percentiles of the residuals in each bin of SFD reddening. In the upper panel, we use unweighted samples drawn from the individual stellar parameter Markov Chains. In the lower panel, we use reweighted samples, as described in the text. \label{fig:reweighted-SEGUE-comp}}
\end{figure}

In Fig. \ref{fig:reweighted-SEGUE-comp}, we compare reddening samples drawn from our PS1+2MASS-based stellar inferences to samples drawn from the SEGUE-based estimates. which have Gaussian uncertainties. The agreement between the two reddening estimates improves significantly after reweighting our PS1+2MASS-based stellar inferences. The improvement is most pronounced at low reddenings. The improvement is negligible for $\EBV \gtrsim 1 \, \mathrm{mag}$, due to the fact that we allow the individual stars to deviate from the local line-of-sight reddening profile by about 10\%. Unlike the SEGUE-based reddenings, the PS1+2MASS-based reddening inferences are constrained to be non-negative, leading to a negative slope in the residuals near zero reddening.

\section{Comparison with Previous Dust Maps}
\label{sec:comparison}

\citet{Schlafly20142d} compares an earlier version of our 3D dust map with the two-dimensional, far-infrared emission-based SFD \citep{Schlegel1998} and Planck dust maps \citep{PlanckCollaboration2013}. Here, we repeat this comparison using our new 3D dust map, and additionally compare our dust map with previous 3D dust maps, which are also based on stellar photometry.

\subsection{2D Dust Maps}

\begin{figure*}
	\plotone{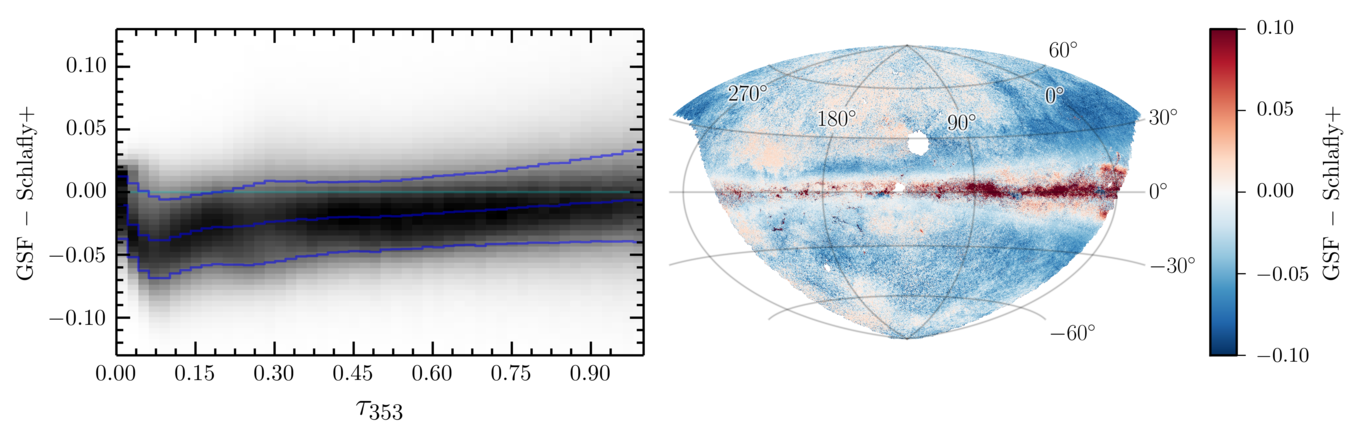}
	\caption{A comparison of our new 3D dust map (``GSF''), integrated to 4.5\kpc, with the 2D map presented in \citet{Schlafly20142d}. The left panel shows the difference between our map and \citet{Schlafly20142d} as a function of a third reddening measure with uncorrelated errors, the Planck $\tau_{353 \, \mathrm{GHz}}$-based dust reddening map. The right panel maps the median residuals between our map and \citet{Schlafly20142d} across the sky. All units are in magnitudes of \EBV. \label{fig:Schlafly-comparison}}
\end{figure*}

It is possible to obtain a 2D dust map from our 3D map by projecting out distance, i.e., by taking the cumulative reddening out to some large distance. Such a map is of particular interest for extragalactic astronomy, where Milky Way dust is essentially a foreground screen to be removed. Several all-sky 2D maps of dust reddening already exist, among them \citet{Burstein1982}, based on H\textsc{i} emission and galaxy number counts, \citet[SFD;][]{Schlegel1998} and two recent reddening maps derived from Planck data \citep{PlanckCollaboration2013}, which model dust optical depth and temperature from far infrared emission, and then calibrate a dust optical depth to reddening relation. Near-infrared stellar colors have also been used to determine dust reddening more directly, as in the NICE/NICER/NICEST family of algorithms \citep[respectively]{Lada1994, Lombardi2001, Lombardi2009}, and \citet{Rowles2009}.

\citet{Schlafly20142d} compares a 2D projection of an earlier version of our 3D dust map with the widely used SFD reddening map, as well as the newer Planck reddening maps. We repeat a number of the same tests for our new 3D dust map.

We begin, however, with a comparison between the map presented in this paper and the map presented in \citet{Schlafly20142d}. The most important differences between the 3D maps used here and in \citet{Schlafly20142d} are addition of near-infrared 2MASS photometry in our newer map, and that we sample here from the full posterior distribution on line-of-sight reddening, rather than finding the maximum-likelihood line-of-sight reddening. Because \citet{Schlafly20142d} requires that each star be detected in $g_{\mathrm{P1}}$, and because we incorporate near-infrared 2MASS photometry alongside PS1 photometry, our map reaches to deeper dust extinctions. This is apparent in the right panel of Fig. \ref{fig:Schlafly-comparison}, where our new map predicts more reddening both in the inner Galactic plane and in dense dust clouds off the plane, such as Orion, Taurus and Perseus.

At reddenings below $\EBV \lesssim 0.08 \, \mathrm{mag}$, our new map predicts significantly less reddening than \citet{Schlafly20142d}, as can be seen in the left panel of Fig. \ref{fig:Schlafly-comparison}. Beyond $\EBV \approx 0.1 \, \mathrm{mag}$, our reddening scales agree very closely, with our new map predicting $\sim \! 4\%$ more reddening. The scatter between the two reddening measures comes to $\sim \! 0.04 \, \mathrm{mag}$, with a maximum median offset of $0.04 \, \mathrm{mag}$ at $\EBV \approx 0.08 \, \mathrm{mag}$, decreasing gradually to an offset of less than $0.01 \, \mathrm{mag}$ between the two measures at $\EBV = 1 \, \mathrm{mag}$. The behavior of the residuals suggests that at very low reddenings, our new 3D dust map prefers essentially zero reddenings too strongly. This may be due to the stronger priors on dust reddening used in the present work, or due to slight differences in our new combined PS1-2MASS stellar locus, versus the PS1 stellar locus used in \citet{Schlafly20142d}.

\begin{figure}
	\plotone{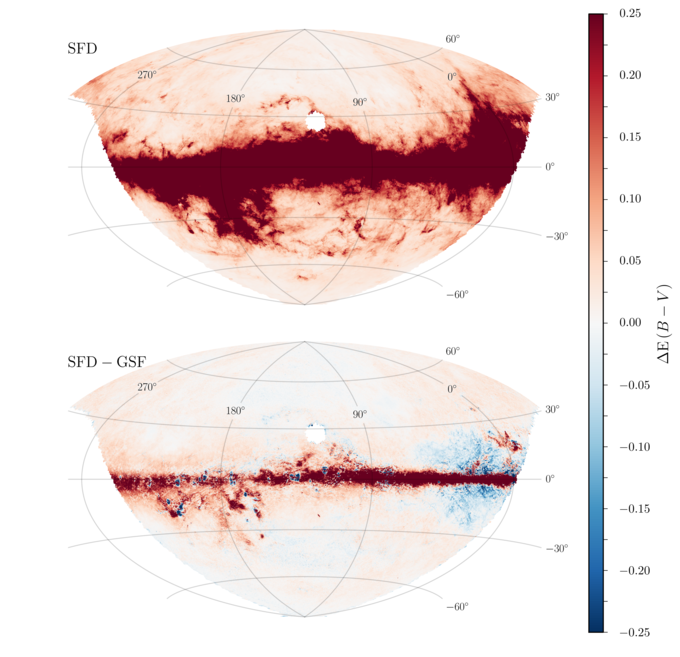}
	\caption{A comparison of the 2D SFD reddening map with our PS1-based map (``GSF''), integrated out to 5 \kpc. The top panel shows the SFD reddening over the footprint of the PS1 survey, clipped to $0.25 \, \mathrm{mag}$, while the bottom panel shows the residuals after subtracting off the median cumulative reddening out to 5 \kpc predicted by our 3D dust map. Both panels use the same color scale. \label{fig:SFD-comparison}}
\end{figure}

\begin{figure*}
	\plottwo{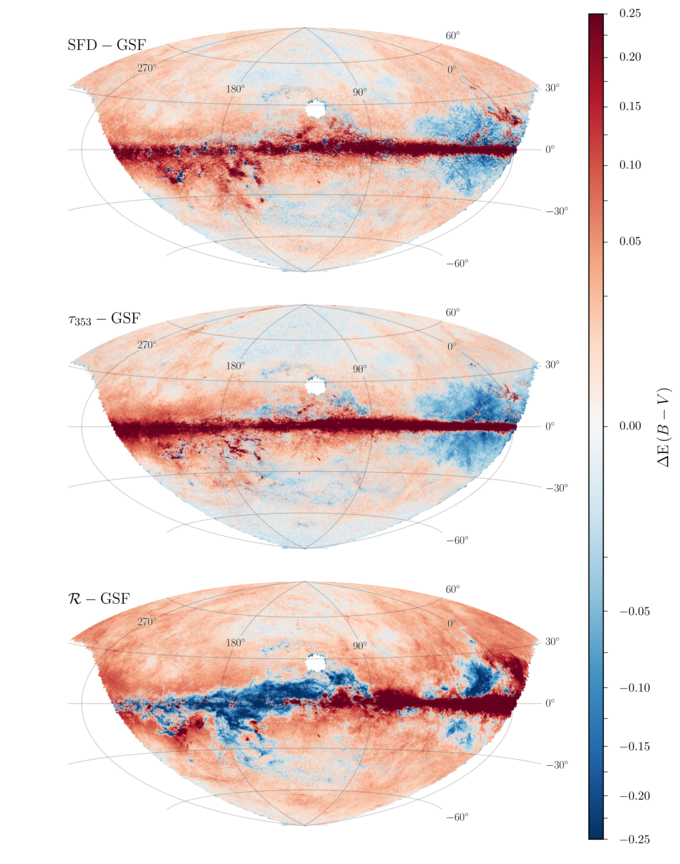}{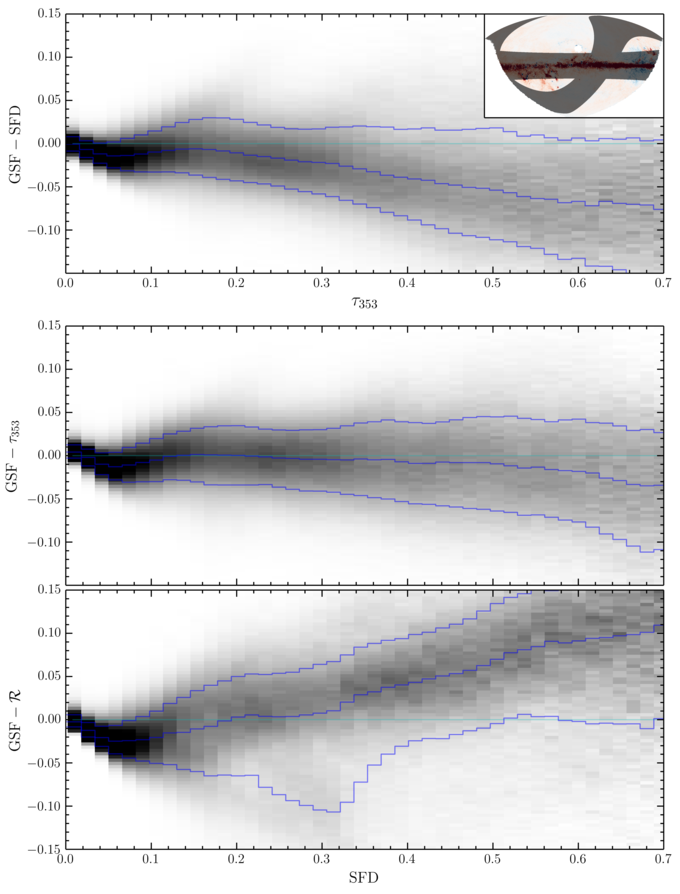}
	\caption{A comparison of our PS1-based reddening to 5 \kpc (denoted by ``GSF'') with SFD, the Planck $\tau_{353 \, \mathrm{GHz}}$-based reddening, and the Planck radiance-based map (denoted by $\mathcal{R}$). The left panels map the residuals (between each map and the median integrated reddening in our map) on a square-root stretch. The panels on the right show the histogram of the residuals (between each emission-based map and posterior samples from our map) as a function of $\EBV$. In the upper right panel, we put the $\tau_{353 \, \mathrm{GHz}}$-based reddening estimate, in magnitudes, along the $x$-axis, since its errors should be largely uncorrelated with the errors in the PS1 and SFD reddening estimates. In the bottom two panels on the right, we use the SFD reddening estimate, in units of magnitudes, as our proxy for reddening, likewise because its errors should be uncorrelated with those of the quantities along the $y$-axis. An inset in the top-right panel shows the regions that are masked in this analysis. The detailed behavior of the residuals, particularly at large reddenings, depends on which regions are masked, indicating that there are systematic differences in the residuals between our reddening map and emission-based map in different regions of the sky. \label{fig:SFD-t353-R-comparison}}
\end{figure*}

Next, we compare our inferred cumulative reddening out to 5 \kpc with the SFD map, the Planck $\tau_{353 \, \mathrm{GHz}}$-based reddening (hereafter, denoted simply as $\tau_{353}$, with units of magnitudes of \EBV), and the Planck radiance-based map (hereafter denoted $\mathcal{R}$, likewise with units of magnitudes of \EBV). All three of these maps are derived by modeling dust emission, and should therefore have different types of systematic errors than our stellar reddening-based dust map.

The upper panel of Fig. \ref{fig:SFD-comparison} shows the SFD reddening across the footprint of our map, while the lower panel shows the residuals after subtracting off our 5 \kpc map. Off the Galactic plane, the residuals are small, while larger residuals are found in the plane of the Galaxy, where there is significant dust past 5 \kpc, and where PS1 does not necessarily detect stars beyond all of the dust. Near the Galactic center, one noticeable feature is a blue halo, where we infer more dust than SFD. The residuals are correlated with dust reddening in this area, indicating a scale offset between the two maps, rather than a constant offset.

The ``blue halo'' is again apparent if we compare our map to the the Planck $\tau_{353}$ map. In Fig. \ref{fig:SFD-t353-R-comparison}, we compare our inferred reddening at 5 \kpc with SFD, $\tau_{353}$, and $\mathcal{R}$. The left panels map the residuals across the PS1 survey footprint on a square-root stretch, emphasizing small-amplitude differences. In the left panels, the Planck-based maps have been scaled by a constant factor to match our inferred PS1 reddening for $\left| b \right| > 20^{\circ}$.

As the ``blue halo'' occurs in the direction of the nearby Aquila Rift, it is possible that the cloud has anomalous dust properties. For example, the grain size distribution or composition may vary, so that $R_{V} = 3.1$ is not a good assumption in the Aquila Rift. Another possibile explanation for the ``blue halo'' is that our stellar inferences may be systematically biased towards greater reddenings in this direction due to limitations in our Galactic model. Our priors do not, for example, include a radial metallicity gradient in the disk components of the Galaxy, which could lead to biased metallicity estimates for stars towards the Galactic center. We leave this question for future investigation.

In the right panels of Fig. \ref{fig:SFD-t353-R-comparison}, we plot the residuals of our inferred reddening to 5\kpc with SFD and the Planck emission-based maps as a function of reddening. Here, we do not scale the Planck maps by any factor to bring them into alignment with our map. To conduct this comparison, we compare the emission-based maps to multiple random realizations drawn from the posterior on reddening to 5 \kpc from our 3D dust map. We restrict our comparison to high-Galactic-latitude regions ($\left| b \right| > 30^{\circ}$), and cut out the ecliptic plane ($\left| \beta \right| < 20^{\circ}$), where imperfectly subtracted Zodiacal light might contaminate the emission-based reddening maps. When comparing our PS1-based reddening with SFD, we place the Planck $\tau_{353 \, \mathrm{GHz}}$-based reddening on the $x$-axis, as its errors are uncorrelated with both maps on the $y$-axis. When displaying the residuals between our PS1-based reddening and the Planck reddening maps, we place SFD along the $x$-axis for the same reason.

For reddenings above $\EBV \gtrsim 0.05 \, \mathrm{mag}$, we see broadly similar residuals as found in \citet{Schlafly20142d}, with with different behaviors above and below $\EBV \approx 0.15 \, \mathrm{mag}$. Above $\EBV \approx 0.15 \, \mathrm{mag}$, our map predicts about 10\% less reddening than SFD, but is in good agreement with $\tau_{353}$. As in \citet{Schlafly20142d}, we find an overall difference in scale between our PS1-based map and the Planck $\mathcal{R}$-based map. For $\EBV \lesssim 0.05 \, \mathrm{mag}$, we see the same residual between our map and the emission-based maps as found between our map and \citet{Schlafly20142d}. For these small reddenings, our map favors essentially zero reddening too heavily.

The exact behavior of the residuals depends on which regions of the sky are masked in the analysis, indicating that there are spatially-dependent systematic differences in the residuals between our reddening map and emission-based maps. However, the essential features remain the same, with different slopes in the residuals below and above $\EBV \approx 0.15 \, \mathrm{mag}$, and our map favoring lower reddening below $\EBV \approx 0.05 \, \mathrm{mag}$.

\subsection{Marshall et al. (2006)}

\citet{Marshall2006a} developed a method to determine the reddening-distance relation along individual lines of sight by comparing 2MASS $J - K_{s}$ stellar colors to those of simulated catalogs based on the Besan\c{c}on model of the Galaxy \citep{Robin2003}. \citet{Marshall2006a} then applied this method to a regular grid of sightlines separated, by $15^{\prime}$ covering the region $\left| \ell \right| < 100^{\circ}$, $\left| b \right| < 10^{\circ}$. The result is a 3D map of reddening in the inner Galaxy, extending to a maximum extinction of $\sim 1.4 - 3.75 \, \mathrm{mag}$ in the 2MASS $K_{\mathrm{s}}$ band (equivalent to $\sim 4.5 - 12 \, \mathrm{mag}$ in $\EBV$), and a maximum distance of $\sim 7 \kpc$. Because \citet{Marshall2006a} use only giants in their analysis, the dust map they produce has little information in the nearest kiloparsec. We will refer to this map as the ``Marshall map.''

\begin{figure*}
	\plotone{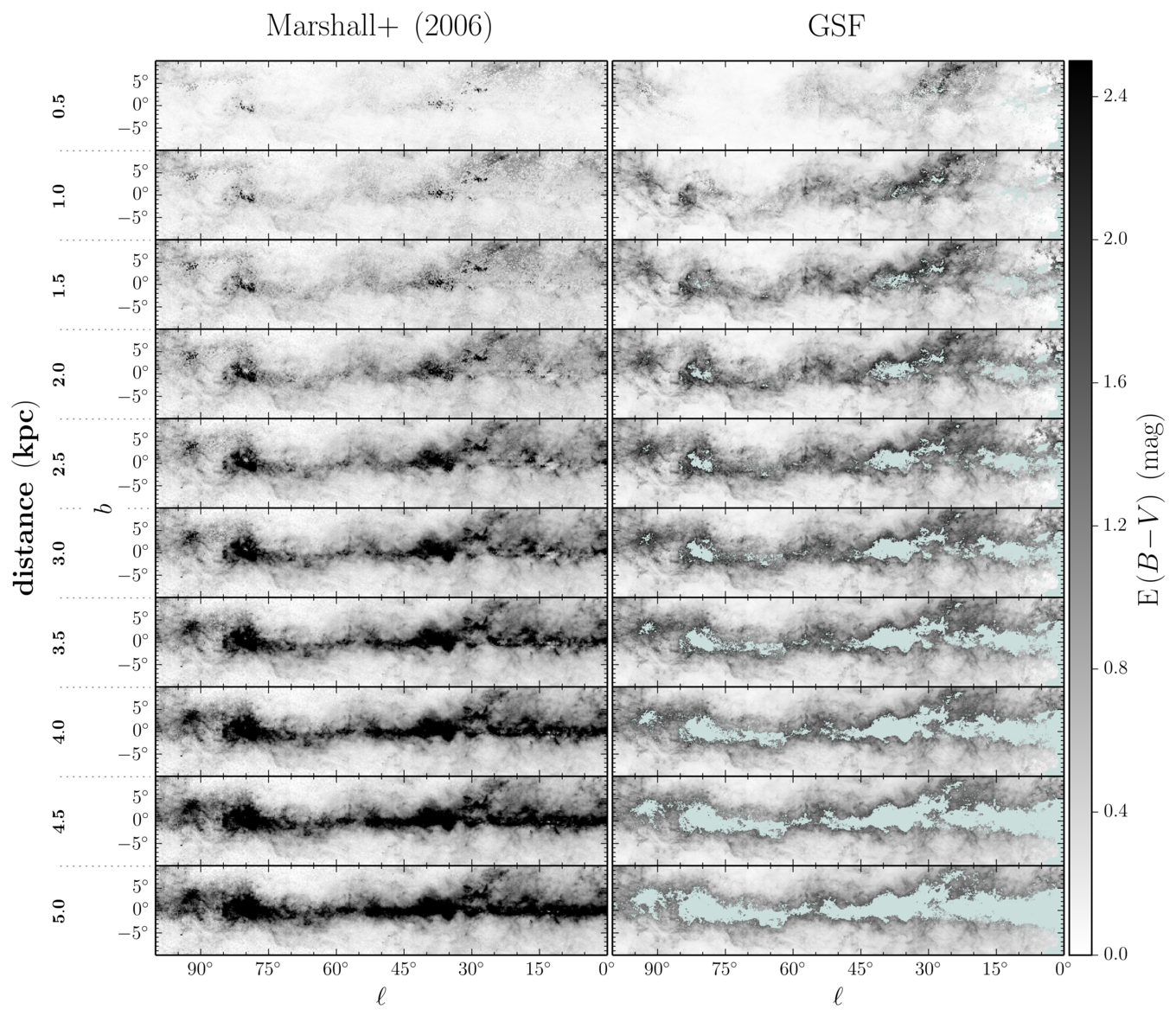}
	\caption{A comparison of the median cumulative reddening out to increasing distances in the Marshall map (left panel) and our map (right panel). Regions beyond the maximum reliable distance in our map are masked out in blue in the right panel. At large distances, the two maps agree qualitatively, with the masked regions in our map corresponding to the most heavily obscured regions in the \citet{Marshall2006a} map. The Marshall map has greater depth, but lower angular resolution, and lower distance resolution in the nearest two to three kiloparsecs. \label{fig:Marshall-comp-cumulative}}
\end{figure*}

\begin{figure*}
	\plotone{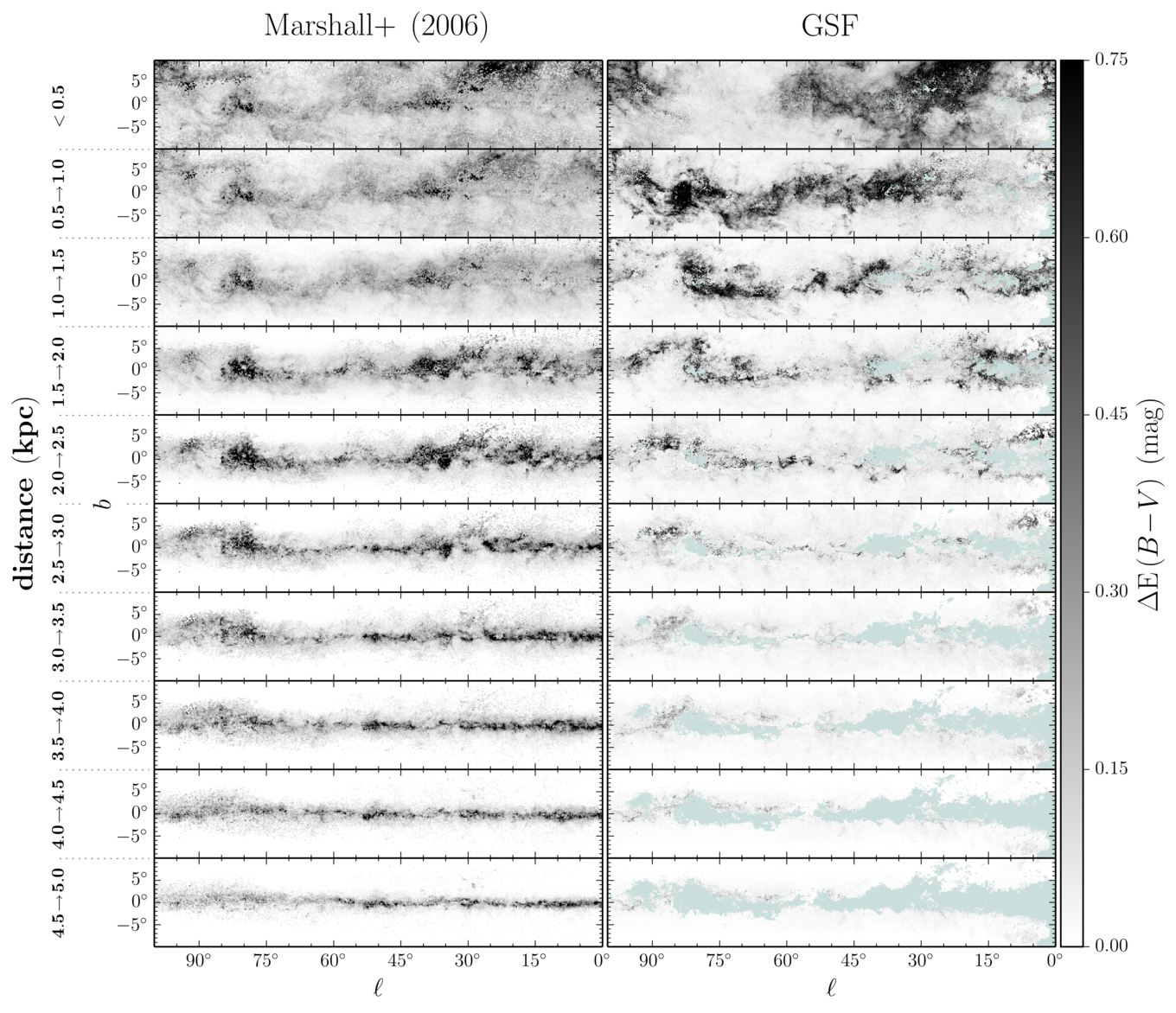}
	\caption{A comparison of the median differential reddening in bins of increasing distance in the Marshall map (left panel) and our map (right panel). In the nearest two to three kiloparsecs, we achieve much better distance resolution, as evidenced by the differentiated structures visible in successive distance bins. \label{fig:Marshall-comp-diff}}
\end{figure*}

\begin{figure*}
	\plotone{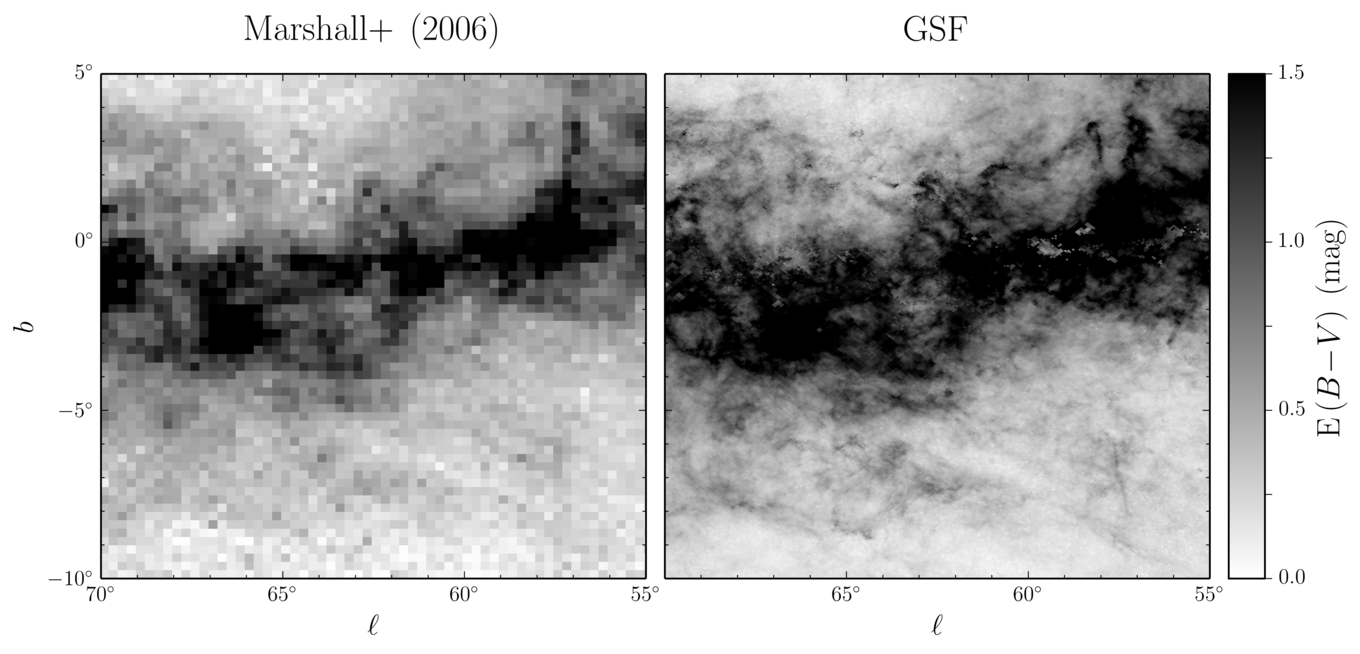}
	\caption{A zoom-in of median cumulative reddening to $3 \kpc$, showing the difference in angular resolution between the Marshall 3D reddening map (left panel) and our map (right panel), as well as the lower noise of the latter. The Marshall dust map has an anuglar resolution of $15^{\prime}$, while our dust map has a typical angular resolution of $3.4^{\prime}$ in the region of the Galactic plane displayed above. The reliability mask has not been applied our map in this figure. \label{fig:Marshall-comp-detail}}
\end{figure*}

Our dust map overlaps with the Marshall map in the approximate region $0^{\circ} < \ell < 100^{\circ}$, $\left| b \right| < 10^{\circ}$. Fig. \ref{fig:Marshall-comp-cumulative} shows the cumulative reddening at increasing distances in both the Marshall map and our 3D dust map. When converting from extinction to reddening, we assume $A_{K \mathrm{s}} = 0.320 \times \EBV$, as calculated by \citet{Yuan2013} for a 7000~K source spectrum at $\EBV = 0.4 \, \mathrm{mag}$, using the \citet{Cardelli1989} reddening law and assuming $R_{V} = 3.1$. We mask our map beyond our predicted maximum reliable distance, as determined in \S\ref{sec:reliable-dists}. In the large-distance limit, our maps show good qualitative agreement outside of the masked areas. Fig. \ref{fig:Marshall-comp-diff} shows the differential reddening in the two maps in bins of increasing distance.

In the nearest two to three kiloparsecs, our map shows clearly differentiated structures at discrete distances that are spread over several distance bins in the Marshall dust map. For example, the Cygnus rift (located at $\ell \sim 80^{\circ}$, $b \sim 0^{\circ}$) appears clearly in our map in the distance bin spanning $0.5 - 1 \kpc$, while in the Marshall map, it is spread over all the distance bins closer than $\sim 3 \kpc$, and is therefore not cleanly separated from superimposed dust structures at greater distances. The greater distance resolution of our map at nearby distances is most likely due to the fact that we use both main-sequence stars and giants, while \citet{Marshall2006a} relies solely on giants, which are saturated nearby and form a larger fraction of the observable stellar population at greater distances.

In addition to better distance resolution in the first few kiloparsecs, the greater source density of PS1 relative to 2MASS allows us to achieve better angular resolution than the Marshall map. Fig. \ref{fig:Marshall-comp-detail} demonstrates the difference in angular resolution between the two maps. In most of the region of overlap between our reddening map and the Marshall map, we achieve an angular resolution of $3.4^{\prime}$, as opposed to the constant $15^{\prime}$ resolution of the Marshall map. This allows us to resolve detailed filamentary structure not seen in the latter map. As dust reddening can vary significantly on small angular scales, this increased angular resolution will be important in practice for correctly de-reddening extinguished sources in regions with complex dust structure.

Deep in the plane of the Galaxy, where high extinction reduces source counts, our finer angular resolution limits the distance to which our map can trace dust, compared to the Marshall map. Although the PS1 $3\pi$ survey is deeper than 2MASS, the 2MASS passbands are less affected by dust extinction, and the advantage of PS1 decreases in regions of high extinction. In such regions, the greater depth of PS1 does not fully compensate for our smaller pixels, limiting the depth to which we trace dust deep in the Galactic plane. The fact that our map derives its most accurate reddening information from main-sequence stars also limits the maximum extinction to which it is reliable.

\subsection{Lallement et al. (2013)}

Combining distance and reddening estimates for $\sim \! 23,000$ stars with the assumption of spatial correlation in dust density, \citet{Lallement2013} infer reddening in 3D out to a distance of $\sim \! 800 \, \mathrm{pc}$ from the Sun. We find close morphological agreement between our 3D dust map and that of \citet{Lallement2013}, with some differences which are worth taking note of.

\begin{figure*}
	\plottwo{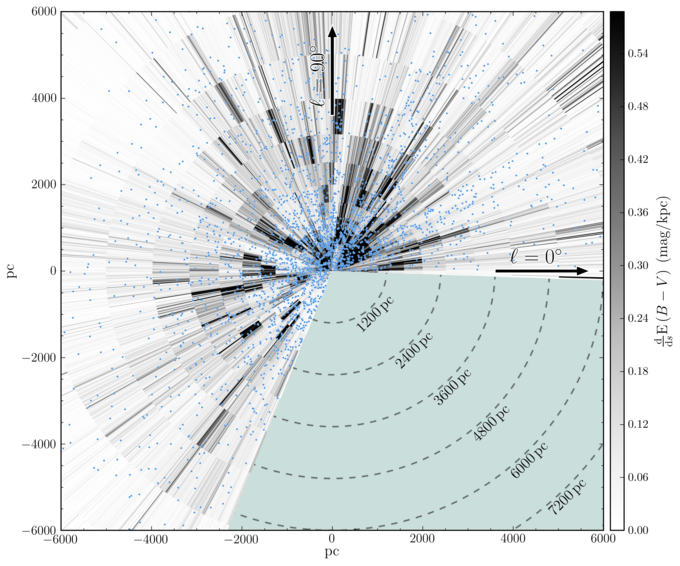}{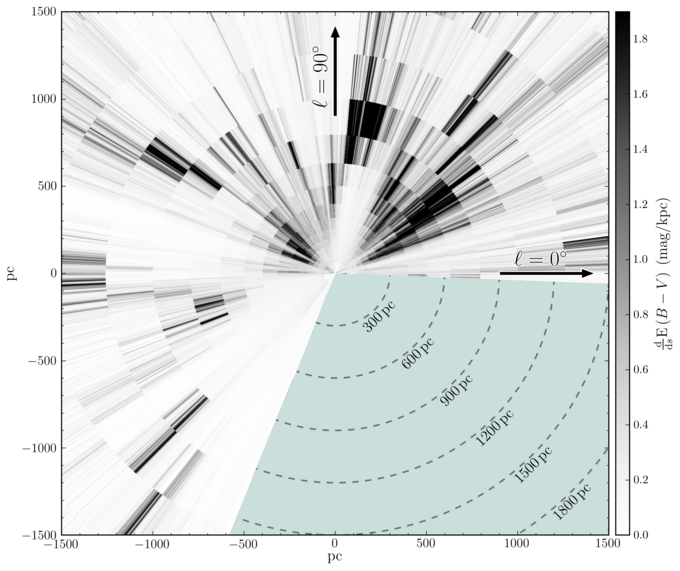}
	\caption{Dust density in a $10 \mathrm{pc}$-thick slice lying $25 \, \mathrm{pc}$ above the Galactic plane (i.e., level with the Sun), with positions of stars within this slice overplotted in the left panel. In detail, we show the median (over multiple realizations of the 3D dust map) of the reddening column density along each sightline passing through the slice (i.e., into the page). The Sun is at the origin of each panel, with the right panel being a zoomed-in version of the left (without the overplotted stars). Only every 1000th star has been plotted. The dust map is reliable out to distances at which the stellar density becomes too small to trace the dust column. Compare with Fig. C2 of \citet{Lallement2013}, in which the stellar density is concentrated within $\sim \! 200 \, \mathrm{pc}$ of the Sun. \label{fig:slice-plane-stars}}
\end{figure*}

In Fig. \ref{fig:slice-plane-stars}, we show the distribution of dust and stars in a slice $25 \, \mathrm{pc}$ above the Galactic plane, level with the Sun. We show the median dust density. In order to generate the stellar locations, we draw a sample at random from the improved distance posterior of each star (see \S\ref{sec:reweighted-stellar-samples}), and select only those stars that lie within $5 \mathrm{pc}$ of the chosen plane. For display purposes, we only display one out of every thousand stars.

Just as \citet{Lallement2013}, we find cavities in the dust density in the directions of $\ell = 70^{\circ}$ and $225^{\circ}$. The overall morphology of the dust structure in the right planel of Fig. \ref{fig:slice-plane-stars} matches that of Fig. 1 in \citet{Lallement2013}.

The most obvious difference between our maps and those of \citet{Lallement2013} is the different voxel shapes employed in our work and theirs. \citet{Lallement2013} use small cubic voxels, and assume a spatial correlation function that favors similar dust densities in nearby voxels. This allows them to densely sample the reddening distribution, with voxels that are not directly constrained by stellar reddening measurements being constrained by neighboring voxels. In contrast, we infer the dust distribution in each line of sight separately, without assuming spatial correlations in the dust density, as laid out in \citet{Green2014}. While our map has excellent angular resolution, it has distance bins with a width of about 25\%, giving our voxels their pencil-beam shape.

Although we see roughly the same structures, such as cavities in the reddening distribution centered on $\ell = 70^{\circ}$ and $225^{\circ}$, the distances we derive for a number of clouds is greater than the distances \citet{Lallement2013} find. In particular, while \citet{Lallement2013} place the Cygnus rift between $500$ and $600 \, \mathrm{pc}$, we place it at a distance of $800$ to $1000 \, \mathrm{pc}$.

\section{Accessing the Map}
\label{sec:accessing}

Our 3D dust map can be accessed at \url{http://argonaut.skymaps.info}. The website provides an interface for querying individual lines of sight, as well as the ability to download the entire map and software to read it. We also provide an API through the website, which allows users to query the map remotely with a few lines of code, without the need to download the entire data cube. The data is also accessible at \url{http://dx.doi.org/10.7910/DVN/40C44C}, through the Harvard Dataverse.

\subsection{Data Cube}

The basic data product our map contains is samples of the differential reddening in 31 distance bins, in each of 2.4 million pixels. The data structure is thus
\begin{align*}
	\left( 2.4 \times 10^{6} \ \mathrm{pixels} \right) \times
	\left( 500 \ \mathrm{samples} \right) \times
	\left( 31 \ \mathrm{distance \ bins} \right) \, .
\end{align*}
This allows us to determine the probability density of the cumulative reddening to any distance (within a few kiloparsecs) in the $3 \pi$ steradians covered by the PS1 survey.

We encourage users to use the Markov chain samples of reddening versus distance provided by our interface in their analyses, as the samples contain the full statistical information generated by our method. These samples can be queried using our web API, downloaded as an ASCII table for individual lines of sight, or accessed directly in the complete data cube provided for download.

An additional, and larger, dataset that we produce while generating the 3D dust map is a library of photometrically determined stellar parameters. As described in \citet{Green2014}, we infer the probability distribution of distance modulus, reddening, metallicity and absolute $r_{\mathrm{P1}}$-magnitude for each star. We thus have a second data cube with shape
\begin{align*}
	\left( 8 \times 10^{8} \ \mathrm{stars} \right) \times
	\left( 100 \ \mathrm{samples} \right) \times
	\left( 4 \ \mathrm{parameters} \right) \, .
\end{align*}
In addition, we store quality assurance information for each star, including whether the Markov Chain converged during the fitting procedure, and the Bayesian evidence for the stellar model, which is similar to the $\chi^{2}$ statistic in maximum-likelihood fitting. Point sources with poor evidence are likely either of stellar types not contained in our model, such as very young stars or binary systems, or are not stars (e.g., white dwarfs, quasars, unresolved galaxies).

\section{Conclusions}
\label{sec:conclusions}

We have presented a three-dimensional map of dust reddening covering three quarters of the sky, based on photometric inferences for $\sim \! 800$ million stars with high-quality multiband photometry in Pan-STARRS 1. The map provides a window into the structure of the interstellar medium, revealing detailed structure from the smallest scales in our map, $3.4^{\prime}$, all the way to large cloud complexes spanning many degrees. We provide interfaces to query and download the map at \url{http://argonaut.skymaps.info}.

Projecting our map down to two dimensions, we find good agreement with emission-based dust maps at high Galactic latitudes, where we expect our method to trace the entire dust column. Comparison with the 3D map of nearby dust presented in \citet{Lallement2013} shows the same morphological features. In comparison with the 3D dust map of the inner Galactic plane presented in \citet{Marshall2006a}, our map has greater angular resolution and superior distance resolution within $\sim \! 3 \kpc$. However, due to our reliance primarily on main-sequence stars, as opposed to giants, our map does not penetrate to as great a depth as \citet{Marshall2006a}.

Our dust mapping technique can be extended to take advantage of photometric surveys beyond PS1 and 2MASS. The Dark Energy Survey \citep[DES;][]{DES2005} is surveying $5000 \, \mathrm{deg}^{2}$ of the southern sky, largely complementary to the PS1 footprint, in a similar filter set. The LSST \citep{Ivezic2008a} will provide deep $ugrizy$ photometry for the sky south of $\delta \approx 34.5^{\circ}$. In the nearer future, the ESA Gaia mission \citep{Lindegren1994} will provide multiband photometry, geometric parallax distance measurements and proper motions for one billion stars. Parallax distances, where available, will vastly improve stellar distance estimates, breaking the dwarf-giant degeneracy in particular. Kinematic information from Gaia will provide additional information about stellar distances and types.

A reliable map the 3D distribution of dust is important to studies of stellar populations within the Galaxy, as well as streams and global morphology of the Milk Way. In future work, we will leverage the stellar inferences produced as a by-product of our map to study the morphology of the Galaxy. We expect that the three-dimensional dust map presented here will find many different uses not yet envisioned by the authors.

\section{Acknowledgements}
\label{sec:acknowledgements}

The Pan-STARRS1 Surveys (PS1) have been made possible through contributions of the Institute for Astronomy, the University of Hawaii, the Pan-STARRS Project Office, the Max-Planck Society and its participating institutes, the Max Planck Institute for Astronomy, Heidelberg and the Max Planck Institute for Extraterrestrial Physics, Garching, The Johns Hopkins University, Durham University, the University of Edinburgh, Queen's University Belfast, the Harvard-Smithsonian Center for Astrophysics, the Las Cumbres Observatory Global Telescope Network Incorporated, the National Central University of Taiwan, the Space Telescope Science Institute, the National Aeronautics and Space Administration under Grant No. NNX08AR22G issued through the Planetary Science Division of the NASA Science Mission Directorate, the National Science Foundation under Grant No. AST-1238877, the University of Maryland, and Eotvos Lorand University (ELTE).

This publication makes use of data products from the Two Micron All Sky Survey, which is a joint project of the University of Massachusetts and the Infrared Processing and Analysis Center/California Institute of Technology, funded by the National Aeronautics and Space Administration and the National Science Foundation.

The computations in this paper were run on the Odyssey cluster supported by the FAS Science Division Research Computing Group at Harvard University.

Gregory Green and Douglas Finkbeiner are supported by NSF grant AST-1312891. Edward Schlafly acknowledges funding by Sonderforschungsbereich SFB 881 ``The Milky Way System'' (subproject A3) of the German Research Foundation (DFG). Nicolas Martin gratefully acknowledges the CNRS for support through PICS project PICS06183.

\appendix

\section{Pan-STARRS 1 and 2MASS Stellar Templates}
\label{app:PS1-2MASS-models}

\citet{Green2014} describes how stellar templates are compiled for the PS1 passbands. Briefly, metallicity-independent main-sequence stellar colors were obtained by fitting a stellar locus in color-color space, and metallicity-dependent absolute magnitudes were obtained from the metallicity-dependent photometric parallax relation given in \citet{Ivezic2008}. For the giant branch, linear fits to globular cluster color-magnitude diagrams from \citet{Ivezic2008} were used.

\begin{figure*}
	\plotone{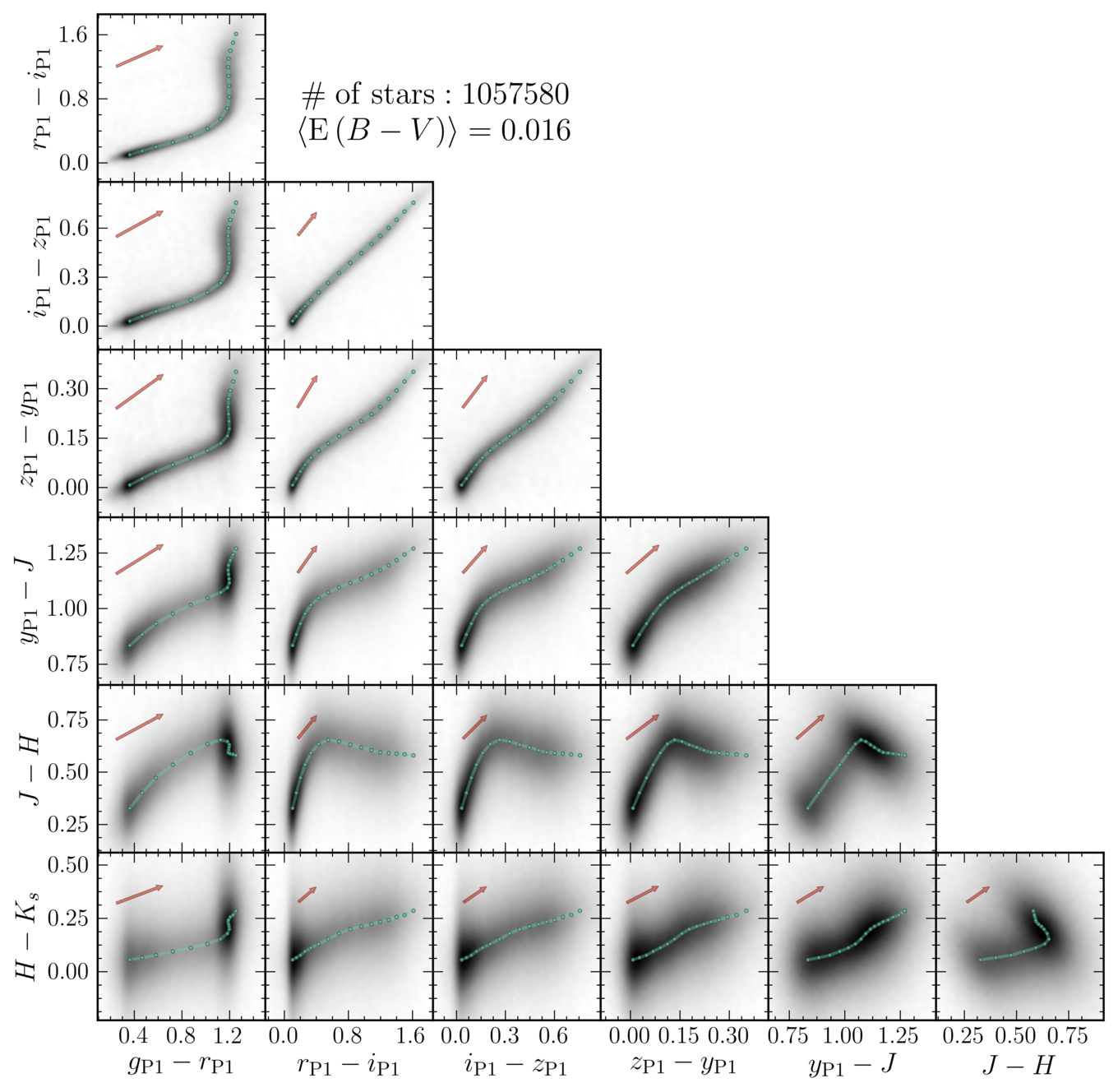}
	\caption{Stellar locus fit in join PS1-2MASS color space. For this fit, $\sim \! 1$ million stars in the vicinity of the North Galactic Pole, with SFD reddening less than $0.1 \, \mathrm{mag}$ in $\EBV$, were used. The fitted stellar locus anchor points are plotted over the density of stars in each color-color projection. A reddening vector with magnitude $0.5 \, \mathrm{mag}$ in $\EBV$ is overplotted for each combination of colors. \label{fig:sl-fit}}
\end{figure*}

In order to include 2MASS photometry in our dataset, we require joint PS1-2MASS stellar templates. We compile these templates using a nearly identical procedure as in \citet{Green2014}. We begin by selecting $\sim \! 1$ million stars with $\EBV_{\mathrm{SFD}} < 0.1 \, \mathrm{mag}$, detections in all PS1 and 2MASS passbands, and photometric errors less than $0.5 \, \mathrm{mag}$ in every passband. The resulting sample has a median reddening of $0.016 \, \mathrm{mag}$ in $\EBV$. After dereddening the photometry, we fit a stellar locus in 7-dimensional color space, using the algorithm laid out in \citet{Newberg1997}. The resulting stellar locus is plotted in Fig. \ref{fig:sl-fit}.

As before, we apply a metallicity-dependent photometric parallax relation to obtain a set of stellar templates, indexed by absolute $r_{\mathrm{P1}}$-magnitude and $\FeH$. In order to obtain templates for the giant branch which incorporate 2MASS, we again use the templates from \citet{Ivezic2008}, replacing stellar colors with our new joint PS1-2MASS stellar locus colors. We use $r_{\mathrm{P1}}-i_{\mathrm{P1}}$ to match each giant template to a color template in our PS1-2MASS stellar locus.

In the 2MASS passbands, we adopt the reddening coefficients $R_{J} = 0.786$, $R_{H} = 0.508$ and $R_{K\mathrm{s}} = 0.320$, calculated by \citet{Yuan2013} for a 7000~K source spectrum at $\EBV = 0.4 \, \mathrm{mag}$, using the \citet{Cardelli1989} reddening law and assuming $R_{V} = 3.1$.

\section{2MASS Selection Function}
\label{app:2MASS-selection}

\begin{figure}
	\plotone{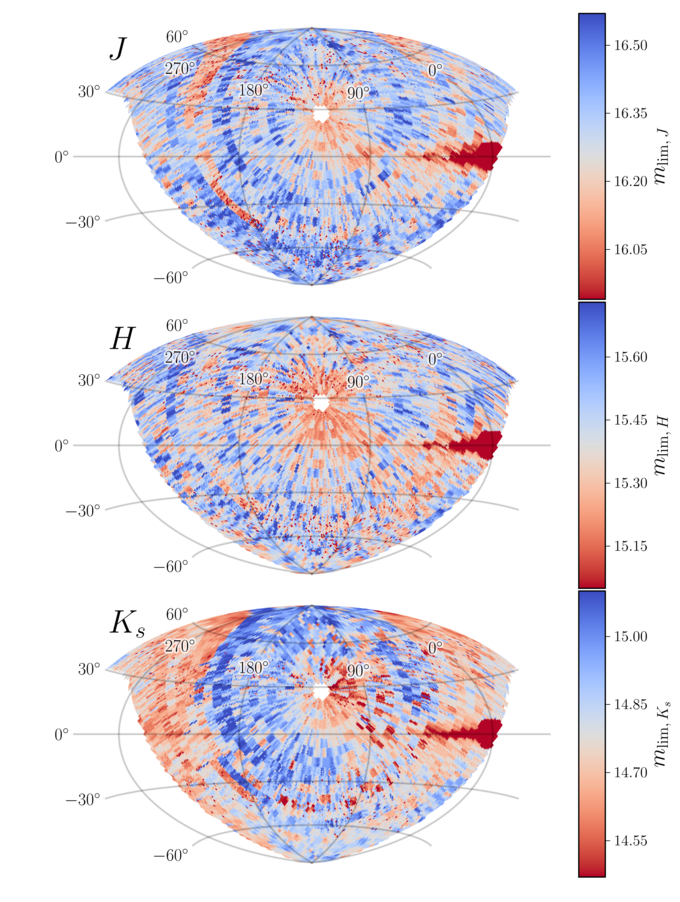}
	\caption{Limiting magnitude of the 2MASS survey in all three passbands, across the PS1 survey footprint. The limiting magnitude is defined as the apparent magnitude at which 50\% of point sources are detected. Effective survey depth is relatively uniform over most of the sky, but falls off precipitously towards the Galactic Center due to source crowding. \label{fig:2MASS-depth}}
\end{figure}

In order to avoid Malmquist bias in photometric inferences, it is necessary to characterize the survey selection function. In \citet{Green2014}, we characterized this as the detection probability in each passband as a function of apparent magnitude and position on the sky. Denote the detection or non-detection of a point source on a particular location on the sky in passband $i$ as a binary variable, $S_{i}$. Denote the intrinsic (or modeled) apparent magnitude of the star in passband $i$ as $m_{\mathrm{mod}, \, i}$. The detection probability is written as
\begin{align}
    p \arg{S_{i} \, | \, m_{\mathrm{mod}, \, i}} \, .
\end{align}
For point-source detections in PS1, we modeled the detection probability based on the sky background, read noise and point-spread function. As the 2MASS point-source catalogs do not contain this information, we model the 2MASS selection function empirically, from the histogram of detections as a function of apparent magnitude across the sky.

There are two ways to approach this problem. The first approach is to construct a full forward model, drawing stellar types, locations and reddenings from our Galactic and stellar priors, generating model photometry for the simulated stars, and applying a trial selection function and photometric errors to obtain a sample of simulated ``observed'' apparent magnitudes. One would then vary the selection function until the histogram of detections versus observed apparent magnitude matched observed histogram in a given region of the sky. This method suffers from its reliance on our relatively crude priors on Galactic reddening, and its sensitivity to any errors in our Galactic and stellar priors.

We therefore opt for a simpler approach. We make the assumption that near the limiting magnitude, the true sky density of objects is a smooth function of apparent magnitude. In small patches of the sky, we locate the turnoff in the histogram of detections, $m_{\mathrm{to}, \, i}$, as a function of apparent magnitude. In the range $-3.5 < m_{i} - m_{\mathrm{to}, \, i} < -0.1$, we model the logarithm of the number of detections in each apparent magnitude bin as a third-order polynomial in magnitude. This is our smooth estimate of the intrinsic sky density of sources as a function of apparent magnitude. For each 2MASS passband, we define the limiting magnitude as the bin in which the observed sky density per unit magnitude falls below 50\% of the estimated true sky density per unit magnitude.

We use this procedure to construct maps of the 2MASS $J$, $H$ and $K_{s}$ limiting magnitudes at two HEALPix resolutions, $\mathtt{nside} = 32$ and 64. In each region of the sky, we adopt the $\mathtt{nside} = 64$ map, unless the given pixel contains fewer than 1000 detections in the given passband, in which case we switch to the lower resolution, $\mathtt{nside} = 32$ map of limiting magnitude. Fig. \ref{fig:2MASS-depth} shows the resulting, multi-resolution map of limiting magnitude in each passband.

\section{Markov Chain ``Swap'' Proposals}
\label{app:swap-proposal}

Because of the way in which we parameterize the line-of-sight distance versus reddening profile, there can be strong anticorrelations between the different parameters. The differential reddening in one distance bin is often strongly anticorrelated with the differential reddening in neighboring distance bins, because transferring some differential reddening from one bin to a neighboring bin keeps the cumulative reddening constant in following distance bins. Similarly, if the data is well fit by a large jump in reddening in one bin, it is often also well fit by a jump in reddening in a neighboring bin.  These strong anticorrelations between the differential reddening in neighboring distance bins can slow down Markov chain convergence when fitting the line-of-sight reddening profile.

\begin{figure}
	\plotone{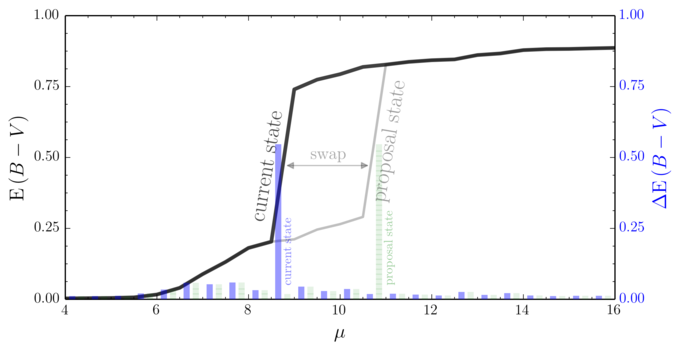}
	\caption{Generation of an MCMC proposal state by the ``swap'' proposal method. The proposal state is identical to the current state, up to a swap in differential reddening in two bins. The black curve shows the cumulative reddening of the current state, while the light gray curve shows the cumulative reddening of the proposal state. The blue bars show the current differential reddening in each distance bin, while the striped green bars show the proposed differential reddening in each distance bin. Note that the two are identical, up to a swap between two distance bins. \label{fig:swap-proposal}}
\end{figure}

We therefore introduce a new type of Markov Chain Monte Carlo (MCMC) proposal step, which we term the ``swap'' proposal. In order to generate a proposal state for the Markov chain, we swap the differential reddening in two distance bins, as shown in Fig. \ref{fig:swap-proposal}.

The swap proposal is symmetric, allowing the usual Metropolis-Hastings acceptance probability to be used. Call the current state $X$ and the proposal state $Y$, and denote the two distance bins that were swapped to generate $Y$ as $i$ and $j$. The probability of proposing $Y$, $Q \arg{X \rightarrow Y}$, is simply the probability that bins $i$ and $j$ are selected to be swapped. As long as the choice of $i$ and $j$ has nothing to do with the current state of the chain, the reverse step, from $Y$ to $X$, is equally likely. That is, $Q \arg{Y \rightarrow X}$ is also simply equal to the probability that distance bins $i$ and $j$ are selected to be swapped. The acceptance probability, $A \arg{X \rightarrow Y}$, is therefore given by
\begin{align}
    A \arg{X \rightarrow Y} &= \mathrm{min} \arg{
    1 , \,\,
    \frac{p \arg{Y}}{p \arg{X}} \, \frac{Q \arg{Y \rightarrow X}}{Q \arg{X \rightarrow Y}}
    } \notag \\
    &= \mathrm{min} \arg{
    1 , \,\,
    \frac{p \arg{Y}}{p \arg{X}}
    } \, ,
\end{align}
where $p \arg{X}$ and $p \arg{Y}$ are the probability densities of $X$ and $Y$, respectively. This is simply the usual Metropolis-Hastings acceptance probability.

The addition of swap proposals, alongside Metropolis-Hastings proposals and affine stretch proposals, allows the Markov chain to mix more quickly. Many distance-reddening curves for the sightline plotted in Fig. \ref{fig:reweighted-stellar-samples}, for example, differ primarily in which distance bin the large jump in reddening occurs. In such a sightline, the addition of swap proposals allows the Markov chain to transition quickly between probable states, greatly improving convergence times.

\section{Scatter in the Line-of-Sight Reddening Profile}
\label{app:los-scatter}

As sketched out in \ref{sec:los-scatter}, we allow each star to deviate slightly from the line-of-sight distance-reddening relation in the pixel. Allowing stars at the same distance to have somewhat different reddenings renders our model more robust to sub-pixel angular variations in dust density. What follows below is a detailed derivation of how marginalization over $\delta_{i}$, the deviation of star $i$ from the average reddening profile in the pixel, translates into a Gaussian smoothing of the stellar probability density function in distance-reddening space, $p \arg{\mu_{i} , E_{i} \, | \, \vec{m}_{i}}$. Beginning with the full posterior density on line-of-sight reddening given our stellar photometry, and including a deviation, $\delta_{i}$, for each star $i$,
\begin{align}
	p \arg{\vec{\alpha} \, | \left\{ \vec{m} \right\}}
	&\propto
    p \arg{\vec{\alpha}} \,
    \prod_{i} \int \! \d \mu_{i} \, \d \vec{\Theta}_{i} \, \d \delta_{i} \,\,
    p \arg{\vec{m}_i \, | \, \mu_{i} , \vec{\Theta}_{i} , \vec{\alpha} , \delta_{i}}
    \notag \\ &\hspace{3cm}
    \times p \arg{\mu_{i} , \vec{\Theta}_{i}, \delta_{i} \, | \, \vec{\alpha}} \, .
\end{align}
Here, $\vec{\Theta}_{i}$ is the type of star $i$. Taking just the integrand, and ignoring the subscript $i$ for the moment,
\begin{align}
    I
    &\equiv p \arg{\vec{m} \, | \, \mu , \vec{\Theta} , \vec{\alpha} , \delta} p \arg{\mu , \vec{\Theta}, \delta \, | \, \vec{\alpha}} \\
    &= p \arg{\vec{m} \, | \, \mu , \vec{\Theta} , \vec{\alpha} , \delta} p \arg{\mu , \vec{\Theta} \, | \, \vec{\alpha}} p \arg{\delta \, | \, \mu , \vec{\Theta} , \vec{\alpha}} \, .
\end{align}
As in \citet{Green2014}, we assume that
\begin{align}
    p \arg{\mu , \vec{\Theta} \, | \, \vec{\alpha}} = p \arg{\mu , \vec{\Theta}} \, ,
\end{align}
with the only complication being the survey completeness limit, which is dealt with explicitly in that paper, but which has no impact on the modification being discussed here. We also assume that
\begin{align}
    p \arg{\delta \, | \, \mu , \vec{\Theta} , \vec{\alpha}} = p \arg{\delta \, | \, \mu , \vec{\alpha}} \, ,
\end{align}
since $\delta$ is a property of the sub-pixel variation in dust density, and should be unrelated to stellar type. Then,
\begin{align}
    I
    &= p \arg{\vec{m} \, | \, \mu , \vec{\Theta} , \vec{\alpha} , \delta} p \arg{\mu , \vec{\Theta}} p \arg{\delta \, | \, \mu , \vec{\alpha}} \, .
\end{align}
The likelihood term (the first term on the right-hand side), can be rewritten as
\begin{align}
    p \arg{\vec{m} \, | \, \mu , \vec{\Theta} , E \arg{\mu ; \vec{\alpha} , \delta}} \, ,
\end{align}
since the modeled apparent magnitude is simply determined by the stellar distance, type and reddening, and the individual stellar reddening is determined by the line-of-sight reddening profile, the stellar distance, and the fractional deviation of the stellar reddening from the local reddening.

By Bayes' Rule, the product
\begin{align}
    p \arg{\vec{m} \, | \, \mu , \vec{\Theta} , E \arg{\mu ; \vec{\alpha} , \delta}} p \arg{\mu , \vec{\Theta}}
\end{align}
is proportional to the posterior density on distance, reddening and stellar type for an individual star, in the presence of a flat prior on reddening. Thus,
\begin{align}
    I
    &\propto p \arg{\mu , \vec{\Theta} , E \, | \, \vec{m}} p \arg{\delta \, | \, \mu , \vec{\alpha}} \, ,
\end{align}
evaluated at $E = \left( 1 + \delta \right) E \arg{\mu ; \vec{\alpha}}$. After marginalizing over stellar type, $\Theta$, we are left with
\begin{align}
    \int \! \d \Theta \, I
    &\propto p \arg{\mu , E} = \left( 1 + \delta \right) E \arg{\mu ; \vec{\alpha} \, | \, \vec{m}} p \arg{\delta \, | \, \mu , \vec{\alpha}} \, ,
\end{align}
Recall that the prior on $\delta$ is a Gaussian centered on zero, with width determined by $E \arg{\mu ; \vec{\alpha}}$. Then,
\begin{align}
    \int \! \d \Theta \, I
    &\propto p \arg{\mu , E = \left( 1 + \delta \right) E \arg{\mu ; \vec{\alpha}} \, | \, \vec{m}} p \arg{\delta \, | \, E \arg{\mu ; \vec{\alpha}}} \, ,
\end{align}
Marginalizing over $\delta$ now leaves us with a smoothed version of the individual stellar posterior density in distance and reddening:
\begin{align}
    \int \! \d \Theta \, \d \delta \, I
    &= \tilde{p} \arg{\mu , E \arg{\mu ; \vec{\alpha}} \, | \, \vec{m}} \, ,
\end{align}
where we have defined the ``smoothed'' individual stellar posterior probability density
\begin{align}
    \tilde{p} \arg{\mu, E}
    &= \int \! \d \delta \, p \arg{\mu , \left( 1 + \delta \right) E \, | \, \vec{m}} p \arg{\delta \, | \, E} \, .
    \label{eqn:posterior-density-smoothing}
\end{align}

In Eq. \eqref{eqn:line-integral-product}, we therefore replace the individual stellar posterior densities with ``smoothed'' posterior densities:
\begin{align}
	p \arg{\vec{\alpha} \, | \left\{ \vec{m} \right\}}
	&\propto p \arg{\vec{\alpha}} \, \prod_{i} \int \! \d \mu_{i} \,\, \tilde{p} \arg{\mu_{i} , E \arg{\mu_{i} ; \vec{\alpha}} | \, \vec{m}_i} \, .
\end{align}

Our method relies on calculating the individual stellar posterior probability densities for all stars along a line of sight first, before sampling from the line-of-sight reddening distribution, as described in detail in \citet{Green2014}. The above derivation shows that this new addition to our method, allowing stars to deviate from the line-of-sight reddening profile, requires an intermediate step, in which the individual stellar posterior densities are smoothed in reddening, according to Eq. \eqref{eqn:posterior-density-smoothing}.

\section{Stellar Sample Reweighting}
\label{app:rw-samples-detailed}

Here, we derive in detail how to reweight the Markov Chain samples resulting from the naive stellar inferences, in order to condition the stellar parameters on the line-of-sight reddening profile. This discussion complements \S\ref{sec:reweighted-stellar-samples}.

Given a fixed line-of-sight reddening profile parameterized by, $E \arg{\mu ; \vec{\alpha}}$, each star is described by a distance $\mu$, stellar type $\vec{\Theta}$, and fractional offset $\delta$ from the reddening profile. The posterior of these parameters, conditioned on the star's photometry and the line-of-sight reddening profile, is given by
\begin{align}
    p \arg{\mu , \vec{\Theta} , \delta \, | \, \vec{m} , \, \vec{\alpha}}
    &\propto p \arg{\vec{m} \, | \, \mu , \vec{\Theta} , \delta , \vec{\alpha}} p \arg{\mu , \vec{\Theta}, \delta \, | \, \vec{\alpha}} \, .
    \label{eqn:rw-posterior-setup}
\end{align}
The second term on the right-hand side can be broken down into two parts,
\begin{align}
    p \arg{\mu , \vec{\Theta}, \delta \, | \, \vec{\alpha}}
    &\propto p \arg{\mu , \vec{\Theta}} p \arg{\delta \, | \, \mu , \vec{\alpha}} \, .
\end{align}
Thus,
\begin{align}
    p \arg{\mu , \vec{\Theta} , \delta \, | \, \vec{m} , \, \vec{\alpha}}
    &\propto p \arg{\vec{m} \, | \, \mu , \vec{\Theta} , \delta , \vec{\alpha}} p \arg{\mu , \vec{\Theta}}
    \notag \\ &\hspace{2.5cm} \times
    p \arg{\delta \, | \, \mu , \vec{\alpha}} \, .
    \label{eqn:rw-posterior}
\end{align}
The two stellar parameterizations, $\left\{ \mu , \vec{\Theta} , \delta , \vec{\alpha} \right\}$ and $\left\{ \mu , \vec{\Theta} , E \right\}$, are equivalent, as the stellar distance modulus, $\mu$, the line-of-sight reddening parameters, $\vec{\alpha}$, and the star's fractional offset, $\delta$, from the line-of-sight reddening are sufficient to determine the stellar reddening, $E$. We would like to reweight the samples we store in the space $\left\{ \mu , \vec{\Theta} , E \right\}$, so that they correspond to the posterior density given above, in Eq. \eqref{eqn:rw-posterior}. As described in \citet{Green2014}, the samples we store in our initial processing are drawn from the posterior density
\begin{align}
    p \arg{\mu , \vec{\Theta} , E \, | \, \vec{m}}
    &\propto p \arg{\vec{m} \, | \, \mu , \vec{\Theta} , E} p \arg{\mu , \vec{\Theta}} p \arg{E} \, ,
\end{align}
with a flat prior on $E$, so that $p \arg{E} = \mathrm{const}$. Transforming to the parameterization $\left\{ \mu , \vec{\Theta} , \delta , \vec{\alpha} \right\}$, we have to transform the flat prior $p \arg{E}$ to an equivalent prior, $p \arg{\delta \, | \, \mu , \vec{\alpha}}$. This prior is given by
\begin{align}
    p \arg{\delta \, | \, \mu , \vec{\alpha}}
    &= p \arg{E} \left. \frac{\partial E}{\partial \delta} \right|_{\mu , \vec{\alpha}}
    \propto \left. \frac{\partial E}{\partial \delta} \right|_{\mu , \vec{\alpha}} \, ,
\end{align}
where we have taken out the constant factor $p \arg{E}$ on the r.h.s. and replaced the equality with a proportionality. The stellar reddening is related to $\mu$, $\vec{\alpha}$ and $\delta$ by Eq. \eqref{eqn:E-delta-relation}. Taking the derivative w.r.t. $\delta$,
\begin{align}
    \left. \frac{\partial E}{\partial \delta} \right|_{\mu , \vec{\alpha}}
    &= \frac{\partial}{\partial \delta} \left[ \left( 1 + \delta_{i} \right) E \arg{\mu_{i} ; \vec{\alpha}} \right]
    = E \arg{\mu_{i} ; \vec{\alpha}} \, .
\end{align}
The flat prior in $E$ thus implies a prior on $\delta$ equal to
\begin{align}
    p \arg{\delta \, | \, \mu , \vec{\alpha}} \propto E \arg{\mu_{i} ; \vec{\alpha}} \, .
\end{align}

In order to sample from Eq. \eqref{eqn:rw-posterior}, we can weight the samples from our initial processing by the ratio of the prior in the new parameterization, where reddening is conditioned on the line-of-sight reddening profile, to the prior used in the initial sampling, where a flat prior on reddening is used. From the above calculation, this ratio is given by
\begin{align}
    \frac{ p_{\mathrm{new}} \arg{\delta \, | \, \mu , \vec{\alpha}} }{ p_{\mathrm{old}} \arg{\delta \, | \, \mu , \vec{\alpha}} }
    &\propto \frac{ p_{\mathrm{new}} \arg{\delta \, | \, \mu , \vec{\alpha}} }{ E \arg{\mu_{i} ; \vec{\alpha}} } \, .
\end{align}
The functional form of $p_{\mathrm{new}} \arg{\delta \, | \, \mu , \vec{\alpha}}$ is given by Eq. \eqref{eqn:scatter-prior}.

\bibliographystyle{apj}
 \newcommand{\noop}[1]{}

\bibliography{bibliography}

\end{document}